\documentclass[a4paper,fleqn,usenatbib]{mnras}

\usepackage[T1]{fontenc}
\usepackage{ae,aecompl}
\usepackage{graphicx}

\title[Beware of Outsiders]{Perturbation of Compact Planetary Systems by Distant Giant Planets}
\author[Hansen] {Bradley M. S. Hansen$^1$\thanks{E-mail:hansen@astro.ucla.edu}\\
$^1$Department of Physics \& Astronomy, University of California Los Angeles, Los Angeles, CA 90095}

\date{submitted August 2016}

\pubyear{2016}

\begin{document}
\label{firstpage}
\pagerange{\pageref{firstpage}--\pageref{lastpage}}
\maketitle

\begin{abstract}

We examine the effect of secular perturbations by giant planets on systems of multiple, lower mass planets orbiting Sun-like stars. We simulate the effects
of forcing both eccentricity and inclination, separately and together. 
We compare our results to the statistics of the observed Kepler data and examine whether these results can be used to explain the observed excess of single
transiting planets. We cannot explain the observed excess by pumping only inclination without driving most systems over the
edge of dynamical instability. Thus, we expect the underlying planetary population for systems with a single transitting planet to contain an
intrinsically low multiplicity population.
We can explain the Kepler statistics and occurrence rates for $R< 2 R_{\oplus}$ planets with a perturber population consistent
with that inferred from radial velocity surveys, but require too many giant planets if we wish to explain all planets with $R < 4 R_{\oplus}$.
These numbers can be brought into agreement if we posit the existence of an equivalent size population of planets below the RV detection
limit (of characteristic mass $\sim 0.1 M_J$). This population would need to be dynamically hot to produce sufficiently strong perturbations
and would leave the imprint of high obliquities amongst the surviving planets. Thus, an extensive sample of obliquities for low mass planets
can help to indicate the presence of such a population.
The histories of our perturbed populations also produce a significant number of planets that are lost by collision with the star and some
that are driven to short orbital periods by the combined action of secular evolution and tidal dissipation. Together or separately these may
provide two channels for the formation of ultra-short period planets as have been observed by Kepler. Some of our simulations also produce
planetary systems with planets that survive in the habitable zone but have no planets interior to them -- much as in the case of our
Solar System. This suggests that such a configuration may not be altogether rare, but may occur around a few percent of FGK stars.

\end{abstract}

\begin{keywords}
celestial mechanics; occultations; planets and satellites: dynamical evolution and stability
\end{keywords}

\section{Introduction}

Over the last decade it has become clear that planetary systems are very common around stars of
mass comparable to or less than the Sun. Observations using both the radial velocity method and the
transit method demonstrate that the frequency of planetary systems  is 
substantial (Howard et al. 2010, 2012; Mayor et al. 2011; Borucki et al. 2011; Youdin 2011; Batalha et al. 2012;
Dressing \& Charbonneau 2013; Christiansen et al. 2015; Petigura, Howard \& Marcy 2013).
Furthermore, it appears that the demographics are dominated by low mass (sub-Jovian) planets 
with semi-major axes $< 1$AU. Indeed, a large fraction of planetary systems appear to be more
compact than our own, often with multiple planets having orbital periods shorter than Mercury.

This has led to an ongoing discussion about the origins of these planets. The original migration
paradigm (Goldreich \& Tremaine 1980; Lin \& Papaloizou 1986; Ward 1997), adapted to explain the
 presence of giant planets on small scales (Lin, Bodenheimer \& Richardson 1996), initially
predicted few small planets with periods less than a year (Ida \& Lin 2008) and resonant chain configurations
for those that did exist (Ida \& Lin 2010). The observational contradiction of these predictions
has spawned alternative models, although the migration paradigm retains a healthy population
of proponents (Rein 2012; Mordasini et al. 2012; Cossou et al. 2014). The late stage assembly of planets in situ 
(Hansen \& Murray 2012, 2013; Chiang \& Laughlin 2013) can broadly reproduce the observed
distribution of planets on small scales, although the mass inventory is still large enough
to imply some level of mass redistribution during the nebula stage, whether it be during the
small particle stage (Boley, Morris \& Ford 2014; Chatterjee \& Tan 2014; Hansen 2014) or as 
planetary embryos (Chatterjee \& Ford 2015, Ogihara, Morbidelli \& Guillot 2015; Boley, Granados Contreras \& Gladman 2016).

Radial velocity surveys also point to the existence of a substantial population of
giant planets at larger ($> 1 AU$) scales, starting with the third exoplanet
system announced (Butler \& Marcy 1996). It is therefore of
interest to understand how giant planets on large scales affect the configuration
of smaller, more compact, planetary systems. While it is quite possible that these
subsystems affect each other in complicated ways during formation, the simple fact of
their mutual existence today 
implies a minimum level of interactions due to
secular oscillations in eccentricity and inclination stemming from  the long-range gravitational
interactions between the components. The characteristic timescales are dictated by the
induced precession which scales $\propto M_p/a_p^{3/2}$, where $a_p$ and $M_p$ are the
semi-major axis and mass of the perturbing planet (e.g. Adams \& Laughlin 2006; Hansen \& Murray 2015; Lai \& Pu 2016). 
This scaling implies that a Jupiter mass planet at 2~AU induces oscillations on the
same timescale as a planet with mass $\sim 4 M_{\oplus}$ with semi-major axis $\sim 0.1$AU.
Thus, there is potential for significant near-resonant gravitational interaction between planetary
systems on large and small scales. Indeed, this is believed to happen in the Solar System, with
 such interactions believed to be responsible
for sculpting features of the asteroid belt (Williams \& Faulkner 1981; Minton \& Malhotra 2011), 
delivering bodies onto Earth-crossing orbits (Froeschle \& Scholl 1986;
Morbidelli et al. 1994),
and contributing to the potential instability of Mercury's orbit (Laskar 1997; Laskar \& Gastineau 2009; 
Batygin \& Laughlin 2009; Lithwick \& Wu 2011).

Thus, a natural question emerges -- given the unexpected nature of the recent planetary discoveries,
what is the potential for dynamical interactions between the compact, low-mass planetary systems
discovered by transit searches and the larger mass, more distant systems probed by radial velocity
surveys? Our first goal in this paper is to examine the likely strength of these interactions and
to identify the mass and separation ranges that allow for interaction.

In fact, there are already potential signatures of such interactions present in the observations
of extrasolar planets today, most notably in the apparent excess of planetary systems showing
 a single transiting planet (see \S~\ref{KSTE}). Thus, a
 second goal of this paper is to identify possible observational signatures of secular
interactions between close and distant planets, and to offer predictions as to how such hypotheses
may be confirmed.

The organisation of this paper is as follows. We will describe
the observational evidence for the existence of external perturbations in \S~\ref{KSTE}. In \S~\ref{IncPert}
we examine the effects of perturbing the inclinations of model planetary systems with a single distant giant planet
 on an inclined, circular orbit. Alternatively, one may perturb the system with a distant giant planet on an
eccentric, coplanar orbit, and we investigate this effect in \S~\ref{EccPert}.
 In \S~\ref{MultiPert} we will
describe how a pair of distant giant planets can increase the strength of perturbations by coupling the
modes of the interior and exterior systems. In \S~\ref{TwoPop} we will examine how  the results 
of the different experiments described in \S~\ref{IncPert}--\S~\ref{MultiPert} can be combined to provide
a range of observable properties.
We also consider, in \S~\ref{LongView}, the possible further evolution of these model planetary systems, due
both to the possibility of long-term chaotic diffusion and the effects of tidal damping.
In \S~\ref{DatComp} we consider possible secondary observables relevant to these results and compare the
results to these in \S~\ref{Disc} before summarising our conclusions in \S~\ref{Conc}.

\section{The Kepler Single Tranet Excess}
\label{KSTE}

The most extensive observational database for model comparison on this question is the Kepler
planet candidate sample (Borucki et al. 2011; Batalha et al. 2013; Burke et al. 2014;
 Rowe et al. 2015; Mullally et al. 2015; Coughlin et al. 2015), which has yielded several thousand confirmed
and candidate planets. A successful theory for planet formation must thus match the observed
properties of this sample, include the distribution of planetary radii, orbital periods and
orbital spacings. Hansen \& Murray (2013) -- hereafter HM13 -- demonstrate that the in situ assembly can match these
observations, but one property of the observed sample was not reproduced, namely the ratio
of systems with one transitting planet (henceforth termed `tranet', as proposed by Tremaine \& Dong 2012) to those with more than one tranet
in the system.

This finding echoed previous analyses. Lissauer et al. (2011a) found that the observed ratios
of systems of different transit multiplicity could  be matched with a single underlying
system of 3--4 planets and an inclination dispersion of a few degrees, as long as one excluded
single tranet systems. This latter class appeared overabundant by as much as 2/3, relative to that
predicted by the best fit models for the higher multiplicities. 
 Other analyses
that attempted to model the underlying population in the face of selection effects 
(Fang \& Margot 2012; Tremaine \& Dong 2012) could similarly only match the population if they allowed
for either a subset with large dispersion in inclination or with a reduced multiplicity.
Ballard \& Johnson (2016) subsequently found that a similar excess of single transit systems is found
around low mass host stars as well, possibly even in greater numbers. Overall the excess of
single tranets appears to be a robust signature of the dynamical history of the observed
exoplanet population.

 Several authors
(Fang \& Margot 2013; Pu \& Wu 2015; Volk \& Gladman 2015) suggest that the highly
multiple Kepler systems are packed close to the dynamically stable limit, and hypothesize
that the lower multiplicity systems are the product of those initial systems that underwent
diminution as the result of dynamical instability. However, Johansen et al. (2012) and
Becker \& Adams (2016) cast doubt on whether such systems, on their own, are capable of
sufficient dynamical self-excitation to actually produce a
sufficient excess of single tranets.
 One obvious route forward is to
consider the effect of perturbations by giant planets on scales $> 1$AU.
We know that giant planets exist on scales $> 1$AU around some fraction of stars, and so the
it is natural to consider that the combination of these and compact planetary systems must exist 
with a non-negligible frequency and may lead to dynamical interaction.

Two broad classes of excitation may serve to dynamically heat our model systems -- if the external perturber(s)
excites primarily planetary inclinations without increasing eccentricities, it may leave the
underlying mass distribution and multiplicity undisturbed, while reducing the frequency with
which tranets are observed. Alternatively, excitation of eccentricity will eventually lead
to the crossing of planetary orbits and result in dynamical instability, reducing the multiplicity
of the planetary system directly. Can one explain the observations with one or both of these models,
or do they require unrealistically strong interactions? Indeed, are they even distinct scenarios,
since excitation of inclination to too large an amplitude will also inevitably excite eccentricities?

\begin{figure}
\includegraphics[width=\columnwidth]{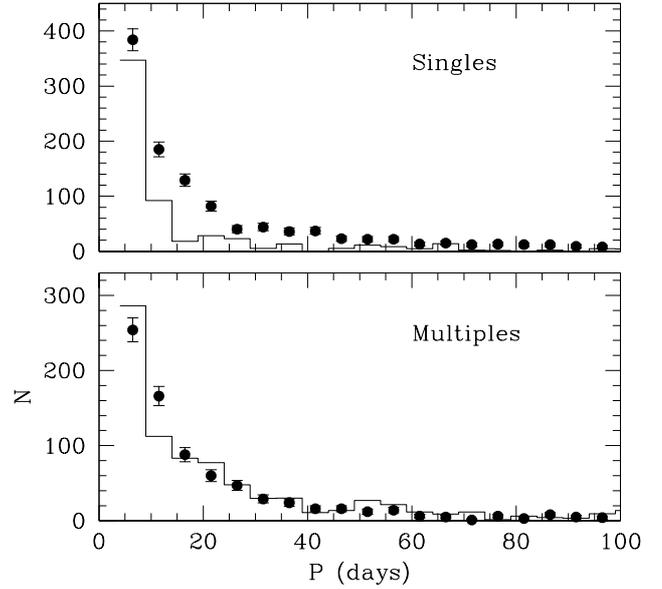}
\caption{The points show the counts of planets with $R < 8 R_{\oplus}$, as a
function of orbital period, for candidates in the
Kepler database orbiting sun-like stars (selected as described in the text). The upper panel shows counts for
stars which show only a single tranet, while the lower panel shows counts for stars with
multiple tranets.
 The solid histogram indicates the
distribution of model planets in our default, unperturbed set of simulations, observed from
random inclinations. The distribution of cases in which only a single tranet is observed is
thus compared to the sample in the upper panel, and those with multiple tranets are shown
in the lower panel. The same scaling has been applied to the model in both panels, and
has been chosen to best match the distribution of multiples in the lower panel.
\label{fig:Pdis_bin}}
\end{figure}

In order to understand the nature of the problem, let us first re-examine the comparison
of the HM13 models with the data. We consider the Kepler Candidate Sample contained in
the NASA Exoplanet Database as of July~5, 2016. We consider two versions of the catalogue.
The latest release is DR24,
 reflecting the analysis of Coughlin et al. (2015). This has the advantage of being the
first to be uniformly selected algorithmically. Unfortunately, Christiansen et al. (2016)
identify some systematic errors in this catalogue at longer periods, which will be of
interest to us later. Thus, we adopt, as our primary source catalogue, the `Cumulative' catalogue
from the archive, although we will quote DR24 results as well, in order to assess the robustness
of our conclusions.
We also restrict ourselves to candidate hosts with 4000~K$<T_{eff}<$7000~K, $4.0<\log g <4.9$ and
Kepler magnitude$<15$, to focus on FGK stars with sufficient signal-to-noise. 
We also restrict our comparisons to those tranets with orbital periods between 4 and 260~days.
The lower period limit is to enable a faithful comparison to the original assembly models of HM13 (which started with an
inner edge to the planetesimal disk of 0.05~AU) and the upper period limit to match the restriction
we place in subsequent sections based on the dynamical stability limit with Jovian mass perturbers
at 1~AU. Figure~\ref{fig:Pdis_bin} shows the period distribution of tranets with $R<8 R_{\oplus}$ for
those in single tranet systems and those in multi-tranet systems. Also shown, as histograms, are
the equivalent samples of model tranets based on observing the simulated systems from HM13 from
randomly oriented lines of sight. These artificial observations make use of the detection efficiency
based on the artificial injection analysis of Christiansen et al. (2016), and discussed in more
detail in \S~\ref{NumExI}.

We see that the observed tranets in single and multiple systems have similar period
distributions, but that the model distribution does not. In the model, the single tranets
are more strongly peaked at short orbital periods, reflecting the fact that the models
show an increase in the inclination dispersion inside 0.1~AU that is amplified by the
increased transit probability at small separations. We can also sum the total number
of tranets and tranet systems in the upper and lower panels and compare them. In the case of the observations,
the ratio of single tranet systems to multi-tranet systems is $3.3 \pm 0.2$, whereas the
models of HM13 predict 1.8. Similarly, if we take the ratio of single tranets to all the tranets found
in multiple systems, the observations yield $1.34 \pm 0.05$ and the models predict 0.71.
The shortfall in these numbers represents the Kepler Single Tranet Excess, or KSTE\footnote{Several authors
use the term `Kepler Dichotomy' to refer to this excess, but we desire a more precise term since
there could be many dichotomies, depending on the context. For those who prefer the more widely
used terminology, a direct substitution throughout this paper will cause no problems.
}. We note that, if we restrict
ourselves to single tranets with period $>10$~days, we reduce the observed count by a factor 0.32,
bringing them below the model values.
 Thus, the excess of single tranets and the difference in the period ratio distribution in
the upper panel of Figure~\ref{fig:Pdis_bin} are closely related issues.

Thus, the question is whether giant planet perturbations can change the properties of the model
systems to better match the observations. In particular, we need to generate more single tranets
on scales $\sim 10$--50~days, as compared to $<10$~days, than are found in the observed systems. As a figure of merit for
comparison, we define $f(50|10)$ to be the ratio of the number of single tranets in the
period range 10--50 days to the number observed between 4--10~days. For the sample defined here,
this is $f(50|10)=1.42 \pm 0.08$, while the unperturbed simulations yield $f(50|10)=0.42\pm 0.03$.
If we calculate the equivalent quantity for the tranets in multiple systems, the observations yield
$1.47 \pm 0.09$ and the simulations $1.37 \pm 0.10$. These demonstrate that there is little difference
in the observations, but a substantial difference in the simulations.

A subsidiary goal of the following analysis is thus to understand what perturbations yield
trends that go in this direction -- increasing the number of single tranets on scales $>$10~days.

\section{Perturbations in Inclination}
\label{IncPert}

A system of earth-class planets on sub-AU scales, with the spacings of observed systems, will interact to produce
secular oscillations of eccentricity and inclination on timescales of $10^3$--$10^5$ years.
These modal interactions manifest themselves as precession of the planetary
orbits. A giant planet on larger scales will also exert an effect, inducing Lagrangian
precession and oscillations in eccentricity and inclination. To quadratic order, the 
evolution of eccentricity and inclination are independent, and so let us consider
each in turn. In particular, we wish to understand if perturbations in inclination
alone can substantially change the transit probability, or does it require such large
perturbations that dynamical instability inevitably follows?

\subsection{Inclination: Increasing the Obliquity or Dispersion?}
\label{LagOi}

To orient our discussion, let us first
  consider a simple
model system of two planets, each of mass $5 M_{\oplus}$, and separated by a factor
of 2 in semi-major axis. We consider this in essence a sub-unit of a multiple planet
system that incorporates the conflict between mutual and external perturbations.
We add to this the influence of a circular orbit giant planet on larger scales, and we ignore the 
backreaction of the interior planets on the outer body. This model problem has a
variety of applications to planetary systems 
(see Boue' \& Fabrycky 2014 or Lai \& Pu 2016 for recent treatments).
For our illustrative purposes, we are satisfied 
to describe
 the behaviour of the planetary inclinations in the classical limit of
low eccentricities and inclinations. To this end 
we define the Delauney variables 
\begin{eqnarray}
p_j & = & i_j \sin \Omega_j \\
q_j & = & i_j \cos \Omega_j
\end{eqnarray}
where $i_j$ are the planetary inclinations and $\Omega_j$ are the longitudes of
the ascending nodes. The secular evolution equations for the inner pair are thus
(e.g. Murray \& Dermott 1999)

\begin{eqnarray}
\dot{p}_1 & = & -\left( B_{12} + B_{13}\right) q_1 + B_{12} q_2 + B_{13} q_3 \nonumber \\
\dot{p}_2 & = & B_{21} q_1 - \left( B_{21} + B_{23} \right) q_2 + B_{23} q_3 \nonumber \\
\dot{q}_1 & = & \left( B_{12} + B_{13}\right) p_1 - B_{12} p_2 - B_{13} p_3 \nonumber \\
\dot{q}_2 & = & -B_{21} p_1 + \left( B_{21} + B_{23} \right) p_2 - B_{23} p_3  \label{sec0}
\end{eqnarray}
where 
\begin{eqnarray}
 B_{12} & = & \frac{1}{4} n_1 \frac{m_2}{m_c}  \alpha_{12}^2 b_{3/2}^{(1)} (\alpha_{12}) \nonumber\\
 B_{13} & = & \frac{1}{4} n_1 \frac{m_3}{m_c}  \alpha_{13}^2 b_{3/2}^{(1)} (\alpha_{13}) \nonumber \\
 B_{21} & = & \frac{1}{4} n_2 \frac{m_1}{m_c}  \alpha_{12} b_{3/2}^{(1)} (\alpha_{12})  \nonumber\\
 B_{23} & = & \frac{1}{4} n_2 \frac{m_3}{m_c}  \alpha_{23}^2 b_{3/2}^{(1)} (\alpha_{23})
\end{eqnarray}
and $\alpha_{ij}=a_i/a_j$, and the $b_j^{(i)}$ are the Laplace coefficients. We assume 
that $p_3$ and $q_3$ are constants, which adds an extra term over the usual expression.
This can easily be converted to the standard form if we transform to a set of new
variables, $u_1 = p_1 - p_2$, $v_1 = q_1 - q_2$, $u_2=p_2 - p_3$, $v_2 = q_2 - q_3$.
This represents the system in terms of the relative orientations of the two inner planets
(index=1) and the outer two planets (index=2). Our evolution equations are then
\begin{eqnarray}
\dot{u}_1 & = & - \left( B_{12} + B_{21} + B_{13} \right) v_1 + \left( B_{23} - B_{13} \right) v_2 \nonumber \\
\dot{u}_2 & = & B_{21} v_1 - B_{23} v_2 \nonumber \\
\dot{v}_1 & = &  \left( B_{12} + B_{21} + B_{13} \right) u_1 - \left( B_{23} - B_{13} \right) u_2 \nonumber \\
\dot{v}_2 & = & -B_{21} u_1 + B_{23} u_2 
\end{eqnarray}
This system can now be solved in the traditional manner to yield two eigenvalues and eigenvectors.
In the limit where the influence of the external perturber goes to zero, this system can be reduced
to
\begin{equation}
 \ddot{u}_1 = - \left( B_{12} + B_{21}\right) u_1 \label{SimHarm}
\end{equation}
leading to a simple sinusoidal oscillation for $u_1$. This is an illustration of the well known fact (e.g. Murray \& Dermott 1999)
that the classical two planet secular problem in inclination yields only a single non-zero eigenvalue because
only the relative inclinations matter. The external perturber provides a reference plane which then
restores the second non-zero eigenvalue. Thus, the behaviour is naturally represented as a combination
of a mode in which the two interior planets oscillate relative to each other and then a mode in which
they oscillate relative to the external perturber. 

\begin{figure}
\includegraphics[width=\columnwidth]{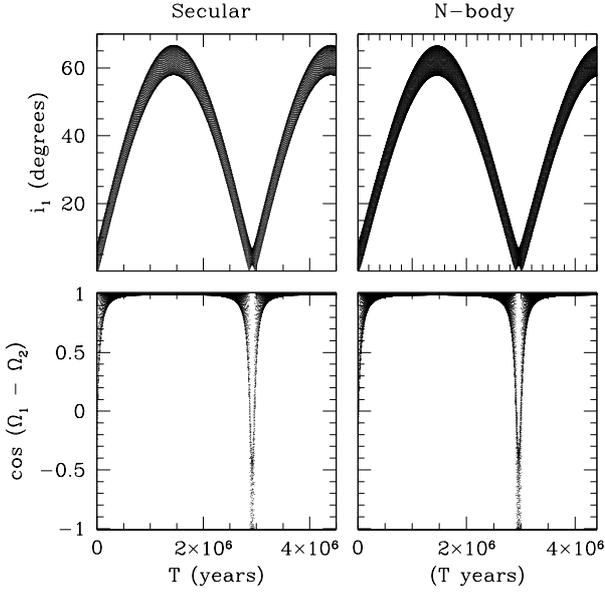}
\caption{The pair of panels on the left show the results of our secular integrations, while
the pair of panels on the right show the integration of the same system with direct N-body integrations.
The upper panels show the evolution of the inclination of the inner planet ($i_1$) while the lower
panels show the evolution of the nodal alignment of the two inner planets (specifically, the cosine of
the difference in their longitudes of ascending nodes).
\label{fig:Ex1}}
\end{figure}

To illustrate the resulting behaviour,
we place our inner pair at 0.1~AU and 0.2~AU, and
place a $1 M_J$ perturber at 5~AU. We give the inner pair initial values of $i_1 = 5^{\circ}$,
$i_2 = 5^{\circ}$ and $\Omega_1 = 75^{\circ}$ and $\Omega_2 = 170^{\circ}$. The outer perturber is
placed on an orbit with $i_3 = 30^{\circ}$ and $\Omega_3 = 0^{\circ}$. This is a little larger than
what one might naively expect to be relevant from a classical calculation, but highlights the intrinsic
behaviour of the model. Figure~\ref{fig:Ex1} shows the resulting secular oscillations. This figure also shows
the same system evolved with the direct N-body code {\em Mercury} (Chambers 1999), demonstrating that
the behaviour is faithfully reproduced, even with the $30^{\circ}$ tilt of the outer planet.

\begin{figure}
\includegraphics[width=\columnwidth]{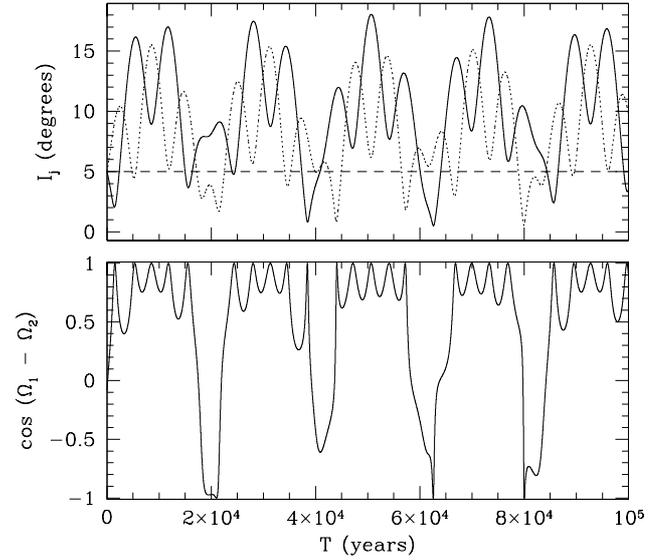}
\caption{The upper panel shows the evolution of the inclinations of the
two inner planets ($i_1$ as a solid line and $i_2$ as a dashed line) when subjected to
a perturber inclined by $5^{\circ}$ (shown as a dashed line). The lower panel again
shows the nodal alignment via $\cos \Omega_1 - \Omega_2$. Deviations of nodal alignment
are associated with periods when one or both of the planetary inclinations drops
substantially below that of the perturber
\label{fig:Ex2}}
\end{figure}

We see that the characteristic behaviour of this system is that the outer planet causes a coherent oscillation
of the inclinations of the inner pair, with a small oscillation in their mutual inclinations. The figures only
show $i_1$ but would look the same if $i_2$ were also plotted, as the amplitude of the difference mode is 
substantially smaller than the mode of their coherent oscillation. We also see that the systems remain  nodally
aligned for most of the time, with only a brief period of misalignment that occurs when the amplitude of the
coherent mode oscillates to low values. The misalignment occurs when the oscillations of the relative mode become
larger than the amplitude of the coherent mode. It is worth also noting that the presence of the constant
terms in equations~(\ref{sec0}) means that the resulting temporal behaviour of the inclinations contains 
power not only at the beat frequency between the two modes (as in the usual case) but also at the two
eigenfrequencies themselves. This can be seen in another example, shown in Figure~\ref{fig:Ex2}, where we repeat
the same calculation as before, but bring the perturber in to $1 AU$ and incline it by only $5^{\circ}$. 
This makes the contributions of the two modes more equal, and yields a more complex behaviour, although it
still retains the same general character.

The inference from this is that the presence of a single distant perturber doesn't necessarily increase
the {\em dispersion} in inclination amongst interior planets, but mostly just tilts the system coherently.
This is has been noted before by several authors (Innanen et al. 1977; Kaib, Raymond \& Duncan 2011;
Bou\'{e} \& Fabrycky 2014; Li et al. 2014; Lai \& Pu 2016).
Thus, a tilted external planet may not change the relative transit probabilities as much as expected.
One question we then need to address is how close a distant perturber has to be to substantially
affect the dispersion and hence the transit frequency.

\begin{figure}
\includegraphics[width=\columnwidth]{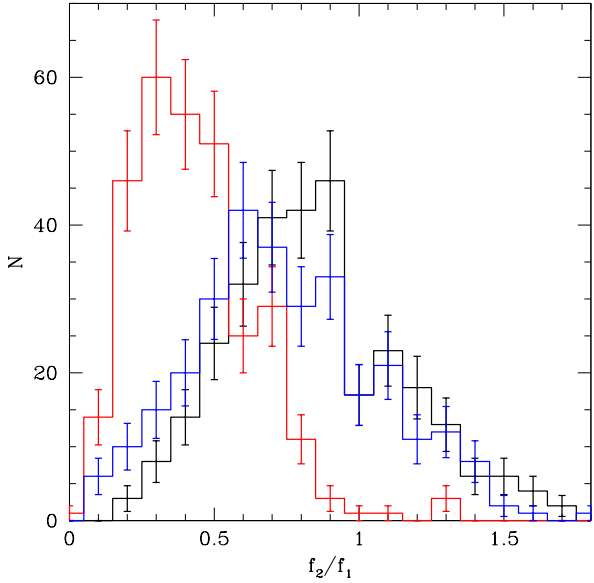}
\caption{The black histogram shows the distribution of $f_2/f_1$ for the case of no perturber, for randomly 
oriented pairs chosen as described in the text. The blue histogram shows the results if we perturb the system with a 
$1 M_J$ planet at 5~AU, with an orbital plane tilted by $5^{\circ}$ relative to the inner pair. If we place the perturber
at 1~AU, we get the red histogram.
\label{fig:Tdis}}
\end{figure}

We can assess this by examining a larger range of systems. We retain the masses and semi-major axes of the inner
pair, but allow for randomly chosen nodal longitudes, and choose inclinations from a Rayleigh distribution with
a dispersion of $1^{\circ}$ (an estimate that fits both inferred distributions -- Fang \& Margot 2012 -- and
assembly simulations -- HM13). We then calculate the resulting secular oscillations in
the case of no perturber, and calculate the observability of the system from a large number of randomly chosen
orientations in order to calculate the probability of observing a single tranet ($f_1$) or two tranets ($f_2$) for the system.
The resulting distribution of $f_2/f_1$ is shown in Figure~\ref{fig:Tdis}. There is a substantial variation depending on the
degree of nodal alignment in a particular system, but the mean value of $f_2/f_1=0.9$, with a dispersion of $0.3$ (although
the actual distribution is asymmetric and biased slightly to lower values).

\begin{figure}
\includegraphics[width=\columnwidth]{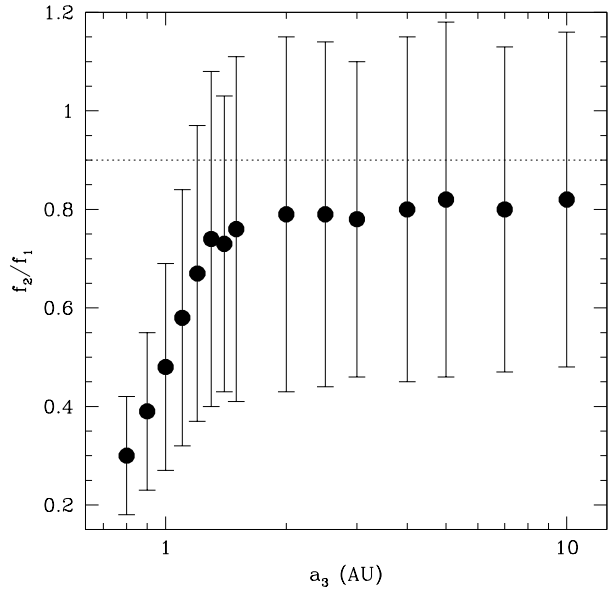}
\caption[Ta]{The points show the mean value of $f_2/f_1$ as a function of the perturber semi-major axis. The perturber is
$1 M_J$ in all cases, tilted by $5^{\circ}$ relative to the reference plane. The error bars indicate the intrinsic dispersion
that results from variations in the inclinations and nodal alignment of the inner pair, as shown in the text. The horizontal
dotted line indicates the value obtained in the case of no perturber.
\label{fig:Ta}}
\end{figure}

If we apply perturbations from a $1 M_J$ planet at 5~AU, with a $5^{\circ}$ inclination, we get the blue histogram shown
in Figure~\ref{fig:Tdis}, and a mean of $f_2/f_1=0.82$, with a dispersion of 0.36. Thus, there is only a small change. If
we move the Jupiter to 1~AU, with the same inclination, then we get the red distribution shown in Figure~\ref{fig:Tdis}. This
has a mean of $f_2/f_1=0.48$ and a dispersion of 0.21. Jupiter mass planets are observed at such distances and so will
notably affect the transit probabilities. However, it is worth noting that these are still well in excess of those observed.
Figure~\ref{fig:Ta} shows how the double/single tranet ratio evolves as one moves the perturber inwards. Indeed, there is little
effect at distances $>2$AU and we only see a substantial effect for $a<1.3$AU. 

\begin{figure}
\includegraphics[width=\columnwidth]{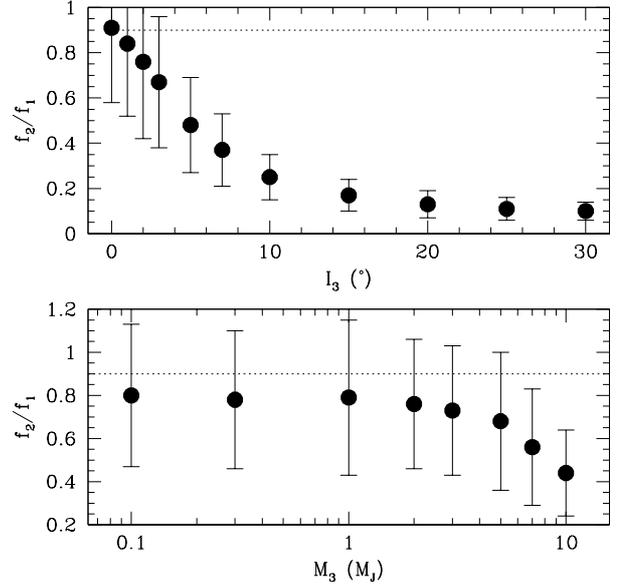}
\caption{The points in the upper panel show the mean value of $f_2/f_1$ as a function of the perturber inclination relative to the
reference plane. The perturber is
$1 M_J$ in all cases, located at 1 AU. The error bars indicate the intrinsic dispersion
that results from variations in the inclinations and nodal alignment of the inner pair, as shown in the text. The
horizontal dotted line indicates the value obtained in the case of no perturber.
The points in the lower panel show the mean value of $f_2/f_1$ as a function of the perturber mass, for a
perturber located at 2~AU and tilted by $5^{\circ}$ to the reference plane.
 The error bars indicate the intrinsic dispersion
that results from variations in the inclinations and nodal alignment of the inner pair, as shown in the text. The
horizontal dotted line indicates the value obtained in the case of no perturber.
\label{fig:Tim}}
\end{figure}

Of course, one expects a greater effect if one tilts the perturber more, although we have shown that the effect is limited.
The upper panel of Figure~\ref{fig:Tim} shows the effect on $f_2/f_1$ as we increase the tilt $i_3$ of a $1 M_J$ planet at 1~AU. We see that we can
indeed get the mean ratio of $f_2/f_1$ under 0.2 as long as we impose tilts larger than $\sim 12^{\circ}$.

We can also make the effects stronger if we increase the mass of the perturber. Higher mass may also mean
a significant influence from more distant perturbers.
The lower panel of Figure~\ref{fig:Tim} shows the effect
of increasing the mass $M_3$ of a perturber at 2~AU, with an inclination of 5$^{\circ}$. We see that we need
a perturber $> 3 M_J$ to start making a difference with these parameters, and still somewhat above the observational
threshold even with $10 M_J$.

These calculations suggest that individual pairs must be perturbed quite strongly before they produce
predominantly single tranet systems. However, real systems often contain more than two planets, with the resulting
increase in the number of available secular modes. Is it possible that an external perturber can couple
more strongly to some modes than others, thereby increasing the inclination dispersion, at least between
different collective modes? Indeed, although
we have noted above that a single planet tends to tilt pairs of planets coherently more than it increases
dispersion, Hansen \& Murray (2015) -- HM15 -- also noted that many of the simulated systems from HM13
broke down into two or more subsets of strongly coupled planets with relatively weak coupling between
the subunits. If only one subunit is tilted relative to another, this might still have some
bearing on the observations. However, too much tilting may also excite eccentricity and thereby incite
dynamical instability. To answer these questions is beyond (semi)-analytic means and requires that
we perform numerical simulations.

\subsection{Numerical Experiments}
\label{NumExI}

We can examine this by considering the model systems produced by HM13. The in situ
assembly integrations in that paper produced a set of model systems that matched most of the properties
of the observed distribution, including distributions in orbital period, period spacing and relative
frequency of multiple transet systems. The most glaring exception was that it did not produce the
excess of single tranets seen in the observations and it is this issue that we wish to address here.
 Thus, we adopt the hypothesis that the results of HM13 serve as a proxy for the underlying high
multiplicity population and examine their sensitivity to a variety of perturbations, seeking to
produce a result that increases the number of single tranets observed.

These are the models used to make the comparison to the data in Figure~\ref{fig:Pdis_bin}.
To do this, we take 50 simulated systems and  observe them at randomly chosen
inclinations to quantify the frequency with which different multiplicities of tranet are observed.
We convert the planetary masses into planetary radii using the mass-radius relation of Seager et al. (2007),
assuming Perovskite equations of state, and increasing the radius by a factor of two relative to that.
This is to account for the fact that many of the planets observed by Kepler appear to have a non-negligible
contribution to the radius from an envelope composed of a lower density material, possibly Hydrogen
(e.g. Lissauer et al. 2013; Weiss et al. 2013; Wu \& Lithwick 2013). A factor of two may be slightly optimistic in this regard, but our goal here is to
examine the influence of perturbations in reducing the observed multiplicity and so we choose to err on
the side of observability. Since the goal of this study is not a detailed inversion of the Kepler data,
but rather a study of the sensitivity of the results to changes in the underlying population, we
 characterise the Kepler detectability of our model planets by assuming high signal-to-noise detections and then
using the results
of an injection and recovery analysis by Christiansen et al., downloaded from the NASA~Exoplanet~Archive (the
details of which can be found in Christiansen et al. 2016), 
to quantify the probability of detection as a function of orbital period and planetary radius.

After observing the original model systems in this fashion, the
 resulting frequency of double to single tranet systems is $f_2/f_1=0.36$. The ratios for
higher multiplicities are $f_3/f_2=0.34$, $f_4/f_3=0.40$ and $f_{4+}/f_4=0.50$. As noted before,
these latter values are consistent with the observations, but the first value is approximately
a factor of two too high. This is the KSTE discussed earlier. In order to calculate a systematic
error bar on these numbers, we randomly select subsets of 25 of this overall set of 50 simulations
and recalculate the tranet frequencies. With a thousand trials, we find that 68\% of trials yield
ratios in the ranges $f_2/f_1=0.355 \pm 0.004$, $f_3/f_2=0.343 \pm 0.005$,
$f_4/f_3 = 0.411 \pm 0.007$ and $f_{4+}/f_4=0.495 \pm 0.015$. Thus, the inferred ratios are robust
with respect to choice of simulated system.

Let us now consider the effect of tilting the system with an external perturber.
We now take the outputs from the original simulations and extend the integrations for an additional $10^7$~years, but with a $1 \rm M_J$ perturber at 1~AU, on a
circular orbit inclined by $10^{\circ}$ relative to the original orbital plane of the unperturbed system.
We designate this numerical Experiment A1. These systems
are once again integrated using the {\it Mercury6} integrator (Chambers 1999) with the
addition of a relativistic precession term. The implementation and testing of this
was discussed in Hansen \& Zink (2015), and we note again here that the short timesteps used in these integrations (12 hours) avoid the
problems sometimes encountered in implementing such effects. The original simulations of HM13 extended out to 1~AU and
so sometimes produced planets that would overlap with the giant planet. In order to avoid planet scattering that
might introduce it's own effects, we removed all original planets with semi-major axis $>0.8$~AU from the initial
conditions before beginning the new integrations.

\begin{figure}
\includegraphics[width=\columnwidth]{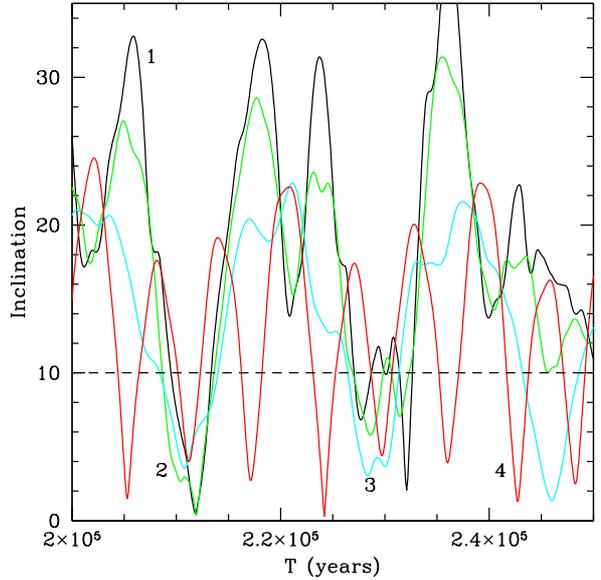}
\caption[ibounce]{The coloured curves show the inclination oscillations
for the innermost four planets of a five planet system perturbed by Jupiter
mass perturber at 1~AU, on a circular orbit tilted by $10^{\circ}$ relative
to the original orbital plane (as indicated by the horizontal dashed line). The fifth most planet oscillates much more
rapidly and was omitted for clarity. The black curve (numbered 1) represents
the innermost planet, while the next innermost (2) is shown in green. The
third closest to the star (3), is shown in cyan and the fourth (4) is shown
in red. This system still exhibits a high frequency of tranet multiplicity when viewed
from randomly chosen angles.
\label{fig:ibounce}}
\end{figure}

\begin{figure}
\includegraphics[width=\columnwidth]{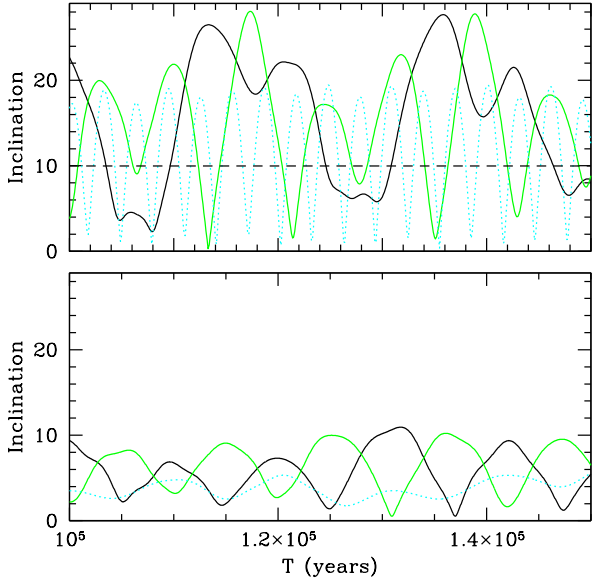}
\caption{The black and green curves in the upper panel show the innermost two of a three planet
system experiencing perturbations from a 1$M_J$ planet at 1~AU on a circular orbit tilted
by $10^{\circ}$ relative to the original orbital plane. The dotted cyan curve shows the outermost
of the trio. The more rapid oscillations for this planet are due to the strong precession induced
by the giant. The inner two oscillate more slowly but still experience substantial perturbations
from the giant. The lower panel shows the same system but without the perturbations from the external
Jupiter. We see the overall amplitudes are smaller and that the third planet is much more in phase
with the second.
\label{fig:ibounce2}}
\end{figure}

Figures~\ref{fig:ibounce} and \ref{fig:ibounce2} show two examples of the behaviour exhibited
in response to the external perturber. In the case of the 5 planet system shown in
Figure~\ref{fig:ibounce}, the inner three planets couple via two modes with substantially
higher frequencies than the external precession induced at these locations, and so they
are largely immune from the effects of the giant planet. The precession of the outermost of the five,
on the other hand, is dominated by the perturber, and is no longer strongly coupled to
the other four. The fourth planet in the system lies in between, coupled to the inner planets
but also experiencing an external forcing of comparable size. The result is the behaviour
shown in Figure~\ref{fig:ibounce}, in which the inclination of the inner three planets track
each other relatively closely, while the inclination of the fourth planet (red curve) moves
in and out of phase with the others. In this system, many multiple tranets are still
seen, and $f_2/f_1 = 0.55$, for the  configuration at the end of the integration, when averaged over all viewing angles.
Of course, this will vary as the planetary inclinations oscillate, and we find that, over the course
of the $5 \times 10^5$~years shown here, the distribution of instantaneous $f_2/f_1 = 0.58 \pm 0.25$.

In the case of the three planet system shown in Figure~\ref{fig:ibounce2}, the perturber
has a stronger effect. In this case, there are only two modes in the absence of the
external perturber. For the third (most distant) of the planets, the resulting precession
is small relative to that from the external perturber and so this one is once again
driven by the giant planet. The other mode couples the two inner planets, and this has
a frequency close to commensurate with the external forcing at the location of the
second planet. The consequence of this is that the inclination oscillations are driven
more strongly and the frequency of tranet multiplicity is reduced substantially. In
the absence of a perturber, this system shows $f_2/f_1=0.43$, but the presence of
the perturber reduces this to $f_2/f_1=0.026$. In part, this is due to the fact that
the second and third planets are responsible for 2/3 of the double tranet observations
in the unperturbed system, and their oscillations are much less correlated when the
third planet is forced by the external perturber.

This, then, is how the tranet multiplicity is reduced across the ensemble as a whole -- some
systems are largely unaffected, while others experience more substantial inflation of their
inclination oscillations and thereby reduce the likelihood of observing a perturbation.
The resulting
frequencies of tranet multiplicity, when sampling the 50 systems as described before, are $f_2/f_1=0.184 \pm 0.002$, $f_3/f_2=0.114 \pm 0.002$, 
$f_4/f_3=0.063 \pm 0.002$ and $f_{4+}/f_4 = 0.417 \pm 0.009$. The value of $f_2/f_1$ is now consistent with what we got for
simplified analysis of planet pairs in \S~\ref{LagOi} and Figure~\ref{fig:Tim}. The factor of two reduction relative to
the unperturbed case also brings it below what is required to match the observations.
 This is
encouraging, although it would imply that nearly all the Kepler systems were subject to a substantial
external perturbation (we will return to this question in \S~\ref{DatComp}). It also now underpredicts
the frequency of higher multiplicity systems, suggesting that this cannot be the entire solution.

The inclinations observed here often reach $\sim 30^{\circ}$, somewhat beyond the classical
expectation. One can wonder whether this starts to pump eccentricities and lead to dynamical
instability? Of the fifty systems integrated, 24 lost at least one planet, either by
collision or ejection. So, some reduction in the number of planets is observed, although the 
level of dynamical instability is not sufficient to drastically alter the true planetary
multiplicity (80\% of surviving systems still contain between 3--6 planets interior to 1~AU). However, the increased mutual inclinations do clearly have some effect
on the tranet multiplicity -- especially on the number of triple and quadruple systems. 
These enhanced inclinations may also
lead to measureable obliquities of planetary orbital planes relative to the stellar spin.
Over the 50 systems observed, the median inclination
for singles is $13^{\circ}$ and for multiples $10^{\circ}$. This is to be compared to the
inclination  distribution of the undisturbed systems, which is well fit by
a Gaussian with dispersion of $4^{\circ}$ for multiples, and a function
of the form $ p(x) \propto x \, {\rm exp}(-0.5 x^2)$
for singles, where $x = i/i_0$ and the inclination dispersion is  $i_0 = 5^{\circ}$. So there  is some tilting
of the overall orbital planes, but probably not enough to be easily discernible
in observations. Not surprisingly, the median obliquities correspond to that
of the perturber.

In terms of the period distribution of single planets, $f(50|10)=0.74 \pm 0.05$, so it is better than
in the unperturbed case, but still somewhat short of the observed distribution. For the
planets in multiple systems, $f(50|10)=1.39$. In principle, we could increase the inclination of the
perturber to reduce $f_2/f_1$ even more, but then it would also reduce the higher multiplicity
statistics. Furthermore, this implies a substantial population of Jovian perturbers, perhaps
too many. Can we achieve the same effects with smaller, distant perturbers?

\subsubsection{Sub-Saturn Perturbers}

  Let us now consider the possibility that our model systems
are coupled to perturbers on larger scales that can still provide sufficient disturbance to reduce the
multiplicities, but are small enough
to evade detection via radial velocities. The current sample of radial velocity planets
trace a curve with amplitude $K \sim 3 m.s^{-1}$, which amounts to a mass $\sim 0.1 M_J$
at 1~AU. Thus, we repeat the above experiment with the same setup but now for a perturber of mass $0.1 M_J$.
We designate this as Experiment~A2.

As one might anticipate, a lower mass perturber with a $10^{\circ}$ inclination yields weaker
perturbations. The multiplicity ratios are $f_2/f_1=0.253\pm 0.003$, $f_3/f_2 = 0.168\pm 0.003$,
$f_4/f_3=0.093\pm 0.003$ and $f_{4+}/f_4 = 0.056\pm 0.003$, which retains the same trend as the $1 \rm M_J$ perturber,
but with a larger ratio of $f_2/f_1$. 

Thus, the reduced perturber mass does reduce the effect and results in multiplicity
ratios closer to the unperturbed model and farther from the observations. However, it is possible
that a larger inclination can yield the same effect with a smaller mass. Thus, we now
repeat our numerical experiment, with mass $0.1 M_J$ but now inclined at $30^{\circ}$ and
$60^{\circ}$. These values might be considered extreme, 
but the last twenty years of exoplanet discovery suggest that perhaps our expectations are
usually too conservative and that we might be better served by Hamlet's famous dictum
''There are more things in heaven and earth, Horatio, than are dreamt of in your philosophy'' (Shakespeare 1623).

Thus, for Experiment~A3, we examine the effects of a $0.1 M_J$ perturber on a circular orbit at 1~AU, but
inclined by 30$^{\circ}$ relative to the original plane of the interior systems. Indeed, the increased
perturbation yields $f_2/f_1=0.080\pm 0.001$, $f_3/f_2=0.035\pm 0.001$ and no quadruple or quintuple tranet systems
at all. The median obliquities are $29^{\circ}$ for singles and $32^{\circ}$ for planets in multi-tranet
systems and $f(50|10)=0.68 \pm 0.05$ for singles. This indicates that, although the tranet multiplicity has been
substantially reduced, the excessive weighting towards short periods remains. On the other hand, these perturbations
leave a large enough obliquity that it may be observable.

\begin{figure}
\includegraphics[width=\columnwidth]{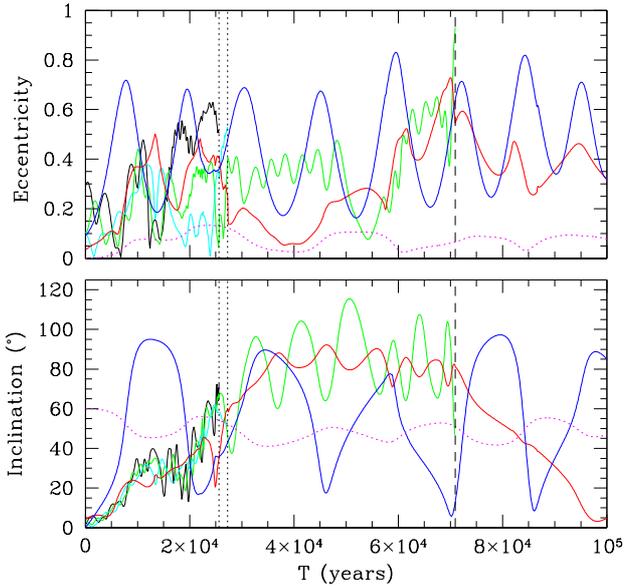}
\caption{The lower(upper) panel shows the evolution of inclination(eccentricity) of
a five planet system operating under the influence of a highly inclined $0.1 M_J$ planet at 1~AU.
As in Figure~\ref{fig:ibounce}, the colours indicate planets ordered from smallest to largest in
semi-major axis as follows: black(1=innermost), green(2), cyan(3), red(4) and blue(5). The dotted
magenta lines indicate the evolution of the perturber, which starts with zero eccentricity and
inclination of $60^{\circ}$. The two vertical dotted lines indicate when planetary collisions
occur and the vertical dashed line signifies the collision of a planet with the star.
\label{fig:Boom2}}
\end{figure}

The larger tilts also result in a lot more dynamical instability. A perturber with  a $10^{\circ}$ tilt
still leaves the bulk of planetary systems numbering from 3--5. With a 30$^{\circ}$ tilt, the predominant
multiplicity of surviving planetary systems is 2 or 3.

We  repeat the experiment with an even larger perturber inclination, of $60^{\circ}$, in order to
examine how much additional attrition is possible (Experiment A4). Indeed, this reduces the surviving multiplicities
to systems with only 1 or 2 surviving planets. This obviously also reduces the statistics
$f_2/f_1=0.0102 \pm 0.0002$ and no triples are observed. More strikingly, this yields an improved value of
$f(50|10)=1.01 \pm 0.07$ for single tranets, twice that of the unperturbed systems, and closer to the observed value.
The reason for this is the dynamically hotter surviving systems have quite a few planets with orbital periods
shorter than the 4 day cutoff we impose. If we include all planets interior to 10 days, this ratio becomes 0.66 again.

 Figure~\ref{fig:Boom2} shows an example of the instability that occurs.
 It shows the same model system as in  Figure~\ref{fig:ibounce}. Once again
the inclinations of the inner three planets grow coherently, but reach such magnitudes that nonlinear
terms lead to eccentricity growth and eventually orbit crossing and scattering. The two innermost
planets collide first and then the third and fourth planets collide. The remaining three planet system
undergoes further interaction until one of the planets is excited to sufficient eccentricity to collide
with the star. After $10^5$~years, there are only two planets remaining and these are subject to continued
dynamical interaction that eventually lead to a collision at 2.9~Myr. This leaves a single planet at
0.22~AU, which undergoes oscillations in eccentricity from $\sim$0--0.2 and inclination from
$\sim$20--60$^{\circ}$ due to continued perturbation from the external planet. Other systems show evolution
in the same general character -- excitation to large enough inclinations prompts eccentricity growth
and ultimately dynamical instability.

This experiment appears to offer a scenario in which one can generate a large number of single
tranet systems with impunity, since the perturbers are hard to detect with radial velocities. However,
the surviving planets are dynamically quite hot, with quite a range of eccentricities and inclinations.
The median inclination of observed tranets from Experiment~A4 is $\sim 59^{\circ}$, so measurements
of stellar obliquity can potentially indicate the presence of such a population. We will examine this
further in \S~\ref{DatComp}.

In summary then, it is possible to reduce the tranet multiplicity by exciting the mutual
inclinations of planetary systems with external planetary perturbers. However, we find that
the level of excitation required to match the observations pushes many systems to the
edge of dynamical instability and beyond. This leads to a reduction in the multiplicity
as well, and thus it is worth examining the alternative pathway to this point -- namely the
excitation of eccentricity directly.

\section{Perturbations in Eccentricity}
\label{EccPert}

In the classical limit, perturbations in eccentricity and inclination are independent and
so, in principle, we can excite orbital eccentricity instead of inclination. Indeed, radial
velocity observations show that giant planets on large scales show a substantial range
in eccentricities which will indeed provide secular pumping for interior planetary systems.
The direct effect of non-zero eccentricities on transit probabilities is modest, but the
pumping of eccentricities in multiple planet systems can drive the system towards dynamical
instability. The merger or loss of planets can dramatically change the system parameters,
and so we must also investigate this avenue.

In principle, one can perform a similar analysis to that of \S~\ref{LagOi} for the case 
of eccentricities, but our goal here is to examine the effects of dynamical instability
and so such an analysis is not as informative. We therefore proceed directly to the numerical experiments.

\subsection{Numerical Experiments}
\label{NumExe}

Let us now repeat our numerical experiments but for an almost coplanar, eccentric perturber.
Unlike our previous experiment, there is substantial observational guidance as to the
strength of the perturbations in this case. Indeed, the circular orbit of our prior inclination experiments is not really
representative of the observed distribution for giant planets on scales $> 1$AU. If we take the Beta distribution of the
observed long period planets from Kipping (2013),
 we find a median eccentricity $\sim 0.2$.

Thus, we perform an experiment (B1) -- integrating the same 50 systems but now with
a perturber of eccentricity $e=0.2$ and semi-major axis 1.25~AU (to yield a periastron
at the same distance as the  semi-major axis in the prior experiments in order to keep the
effects of direct scattering at a similar level). We will again assume $1 M_J$ but
keep the planet almost coplanar (an inclination of $1^{\circ}$ to avoid overly artificial symmetries). We integrate these
systems for $10$~Myr and calculate the odds of transit as before.
 The resulting multiplicity
ratios are $f_2/f_1 = 0.287\pm 0.001$, $f_3/f_2=0.270\pm 0.002$, $f_4/f_3=0.278\pm 0.002$
and $f_{4+}/f_4=0.130\pm 0.002$. Thus, the multiplicity ratios are reduced relative to the
unperturbed case, but not as much as in Experiments A1, A3 and A4, and certainly not 
to the point that they agree with observations. 
 The ratio
$f(50|10)=0.62 \pm 0.04$ for singles, which is higher than the unperturbed systems, but 
still well short of observations.
The value for the multi-tranet systems, $f(50/10)=1.80 \pm 0.11$, is in better agreement.

The manner in which the reduction in multiplicity is achieved is also a little different. If we take the
same five planet system as depicted in Figure~\ref{fig:ibounce}, we find that the inner
four planets of the system are driven to merge into a single object by a series
of collisions. The cause of this is the fact that the increased apsidal precession
of the fifth planet, driven by the giant planet, brings it into secular resonance with
a mode that couples the inner three planets, causing eccentricity pumping (the fifth
planet is approximately the same mass as the inner four combined) of the inner system,
to the point that their orbits cross and they collide. Ultimately this leaves behind
a widely separated two planet system. However, the pair has only a small inclination
dispersion and still yields a relatively high ratio of $f_2/f_1$ (although obviously
no higher multiples)!

In the case of the second system discussed before (and shown in Figure~\ref{fig:ibounce2}),
the multiplicity is also reduced to two, but in this case it is the outer two planets
that collide. The eccentricity of the outermost planet is pumped sufficiently strongly
to cross the orbit of the second planet, leading to a collision. The resulting widely
separated pair also has a sufficiently large dispersion in inclination that it also
exhibits $f_2/f_1=0.04$.

The consequence of this is that the pumping of both eccentricity and inclination can
serve to decrease the multiplicities, both of tranets and of the underlying planets.
 In the case of the tilted, circular $1 M_J$ perturber, 24/50 systems saw some reduction
in the number of planets, whereas the eccentric perturber reduced the multiplicity in
31/50 systems. In total, these 31 systems saw 46 planetary collisions, 6 collisions
with the star, 1 collision with the external perturber and 22 ejections. In this case
80\% of the surviving systems have multiplicities between 2--5, so a slight reduction
relative to Experiment~A1, but still with many multiple planet systems.

As in the case of the inclinations, we can also repeat the experiment with a smaller
mass perturber. In Experiment~B2, we rerun the simulations with the same semi-major axis
and eccentricity, but with a perturber mass of $0.1 M_J$. 
 The resulting systems are much less affected by the perturbations and the resulting
 tranet multiplicities are $f_2/f_1=0.289\pm 0.001$, $f_3/f_2=0.354 \pm 0.002$, $f_4/f_3=0.420 \pm 0.007$
and $f_{4+}/f_4=0.118\pm 0.002$, too large
to match the observations. Furthermore, $f(50|10)=0.61 \pm 0.04$ for the singles, which is similar
to that of the unperturbed systems.
Thus, the reduced perturber mass does reduce the effect and results in multiplicity
ratios closer to the unperturbed model and farther from the observations.

Of course, we noted in \S~\ref{IncPert} that we can compensate for lower masses by
increasing the level of excitation. Thus, in Experiment~B3, we consider the effect of a
0.1~$M_J$ perturber with e=0.5. In order to avoid direct scattering between the perturber
and the unperturbed orbits, we now move the perturber semi-major axis out to 2~AU. The
results are less dramatic than in the inclination case, as $f_2/f_1=0.300 \pm 0.006$, which
is still too high to match observations. The higher multiplicity statistics are reduced
though, with $f_3/f_2=0.197 \pm 0.005$, $f_4/f_3=0.163\pm 0.006$ and $f_{4+}/f_4=0.156\pm 0.008$.
The period distribution is also not markedly affected, with $f(50|10)=0.66 \pm 0.04$.

\begin{figure}
\includegraphics[width=\columnwidth]{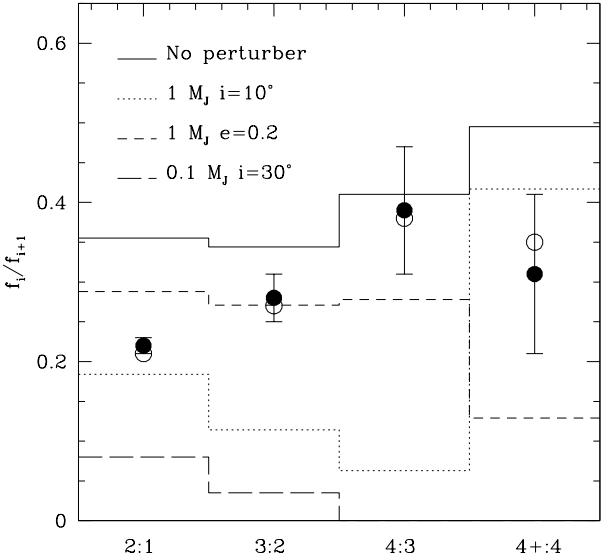}
\caption[Rat3]{The solid points shows the multiplicity ratios of the Kepler candidate
sample calculated from the July 2016 Cumulative sample, with Poisson
error bars, when considering only systems in the period range 4--260~days (the most
accurate comparison to our simulations). The open points are calculated
using the DR24 (Coughlin et al. 2016) candidate sample. The solid histogram is the expectation from our model simulations without
any external perturber, when
observed from random orientations. The addition of a circular Jupiter at 1~AU, include
by 10$^{\circ}$, yields the dotted histogram. If we change the perturber to be coplanar
but with an eccentricity of 0.2 (and move it to semi-major axis 1.25~AU), we get the
short dashed histogram. If we instead perturb the systems with a mass reduced to 0.1~$M_J$,
and inclined by $30^{\circ}$,
we get the long dashed histogram. Thus, moderate eccentricity perturbations preserve
the higher multiplicity ratios, but do not lower the 2:1 ratio enough. Moderate inclination
perturbations reduce the 2:1 ratio sufficiently, but reduce the higher multiplicities  too.
All model populations are subjected to selection effects based on the Christiansen et al. (2016)
injection-recovery tests, as described in the text.
\label{fig:Rat3}}
\end{figure}

Figure~\ref{fig:Rat3} summarises these results, by comparing the multiplicity ratios
from the various experiments to those of the observed population. 

The consequence of these experiments is that the influence of a single moderate perturber
can reduce the multiplicities at a sufficient level to match the observations, but requires
 planetary parameters that should be observationally testable.
If the perturber has a moderate inclination or eccentricity, it must be of sufficient mass to be detectable by radial velocity surveys. 
Alternatively, perturbers smaller than radial velocity limits can still disrupt systems if they have large enough
inclinations. In this case, the surviving planetary systems display substantial obliquities. This suggests it
should be possible to observationally probe such scenarios, and we examine this question
in \S~\ref{DatComp}.

\section{Multiple Perturbers}
\label{MultiPert}

Thus far, we have considered the effects of a single external perturber. In essence, the
simulated systems are conceptually no different from the original unperturbed systems
except that the coupling to the perturber can potentially alter the eigenvalues and increase
their amplitudes to levels which may affect the observability or stability of the system.

However, if the perturbations result from a system of giant planets on larger scales, this
has the potential for qualitatively different evolution. The presence of multiple giant
planets means that they form a gravitationally interacting subsystem of their own, with
its own characteristic secular oscillations. This opens up the possibility of resonance
between the modes of the inner and outer systems, which can drive the oscillatory amplitudes
to even larger amplitudes. Indeed, the most dramatic potential consequences of secular
perturbations in the solar system are not simply the result of Jupiters presence, but rather
its precession and the resonant interaction with the inner terrestrial planets (e.g.
Lithwick \& Wu 2011; Batygin, Morbidelli \& Holman 2015).

\subsection{Conditions for Resonance}

A simple way to represent the effects of multiple external planets is if we impose a precession on the
innermost perturber (such as may be induced by a second giant planet), which amounts to a linear ramp in $\Omega_3$ or $\bar{\omega}_3$, with some
characteristic frequency. The aforementioned model is still well described by
as a system of coupled harmonic oscillators, but now it also features a forced
component, with a specific frequency. Such systems are known to exhibit strong resonant responses when
the forcing frequency corresponds to one of the natural frequencies of the
system. However, how plausible is it that such resonances occur?

\begin{figure}
\includegraphics[width=\columnwidth]{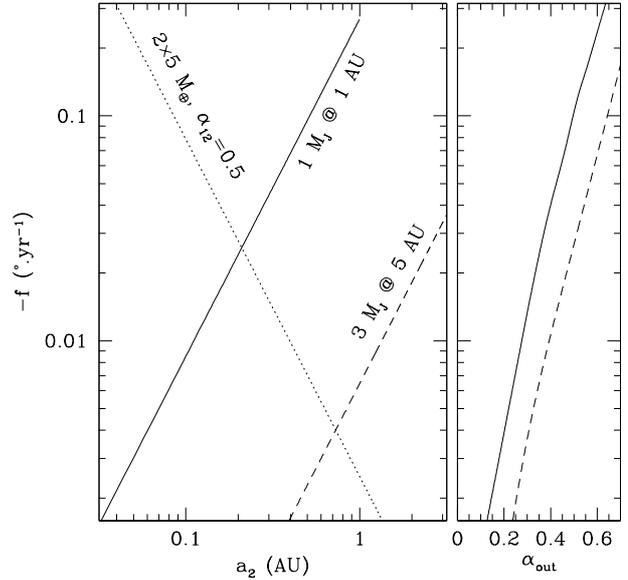}
\caption{In the left hand panel, the dotted line shows the nodal precession frequency
for a pair of planets, each of mass $5 \rm M_{\oplus}$, separated by a ratio of 0.5 in semi-major axis.
The frequency is plotted as a function of the semi-major axis of the outer of the planet pair.
The solid line shows the externally induced precession felt at the same location due to a Jupiter
mass planet at 1~AU. The dashed line shows the precession induced by a $3 M_J$ planet at 5~AU. 
The right hand panel shows the semi-major axis ratio required for a pair of Jupiter mass planets
if they are to match the precession frequency shown in the left-hand panel. The solid and dashed
lines refer to the same curves in the left-hand panel, where the planet is now the innermost of a pair.
\label{fig:Prec}}
\end{figure}

To estimate the degree to which external systems couple, let us consider again 
an idealised system of two planets, much like in \S~\ref{LagOi},
 -- a pair of $5 M_{\oplus}$ planets
seperated by a semi-major axis ratio $\alpha_{12}=a_1/a_2=0.5$, but we now allow $a_2$ to vary. The mutual interactions result
in a secular precession of the nodes with a frequency
\begin{equation}
f_{in} = - 0.079 ^{\circ}.yr^{-1} \left( \frac{a_2}{0.1 \rm AU} \right)^{-3/2}
\end{equation}
If we expect that the precession due to an external planet, located at semi-major axis $a'$, substantially influences the 
properties of the planet pair, then we anticipate that this precession $f_{io}$ should be comparable
to or greater than $f_{in}$. This is given by
\begin{equation}
f_{io} = - 0.0085 ^{\circ}.yr^{-1} \left( \frac{a_2}{0.1 \rm AU} \right)^{3/2} 
\left( \frac{a'}{1 \rm AU} \right)^{-3} \left( \frac{M'}{1 M_J}\right)  \label{io}
\end{equation}
where $M'$ is the mass of the external perturber, and we have assumed that the separation
is large enough to approximate the Laplace coefficient by it's leading order term.
Furthermore, if we wish to drive a resonant interaction with an outer planet pair,
we also require that the mutual precession frequency of the outer pair be comparable,
\begin{equation}
f_{out} = -0.09 ^{\circ}.yr^{-1} \frac{M'}{M_J} \left( \frac{a'}{1 \rm AU} \right)^{-3/2} g(\alpha') \label{outerf}
\end{equation}
where
\begin{equation}
 g(x) = x^{7/4} b_{3/2}^{(1)}(x) \left( 1 + \sqrt{x} \right)
\end{equation}
is a function of $\alpha'$, the semi-major axis ratio of the outer pair. In this case we have assumed
that the two external planets are of equal mass $M'$. Thus, for a given inner pair, we can constrain
the locations of an external planet of given mass such that it has a non-negligible influence and then
constrain the potential locations of a second planet exterior to that which can drive it at an
appropriate frequency to satisfy $f_{out} \sim f_{io}$.

Figure~\ref{fig:Prec} shows these constraints graphically. Examination of this plot suggests that Jupiter-mass
planets at 1~AU will drive precession down to $\sim$ 0.3~AU and that outer pairs with $\alpha'>0.4$ 
(period ratios $< 4$) will have plenty of cross-section to interact with inner pairs
on those scales. As we move the perturbers outwards the influence weakens, and we show a second
example of a system $M' =3 M_J$, $a'=5$AU that can only couple substantially to inner pair 
if  $a_2>0.7$AU, which is starting to approach the limit of our original
simulations. 

This suggests that the likely range of interest for external perturbations is perturbers in the
range 1--5~AU. Also, it suggests that it doesn't require extremely close giant planet pairs to have
interesting interactions with the interior systems, as long as the inner planet is close enough. 
Wider separation pairs will drive resonances closer to the star and closer pairs will provide
excitations on larger scales.

\subsection{Resonant Driving of Inclination}
\label{MultiPerti}

To investigate this effect,
 we now perform Experiment C1, in which we adopt the same setup as before, a $1 M_J$ planet
 with semi-major axis 1.25~AU, but we now add
 a second $1 M_J$ planet at
3.5~AU. The separation is chosen to be
representative of the kinds of giant planet separations identified in giant planet pairs detected
by radial velocities,  when selecting those whose inner member resides further
than 1~AU. We base this on the median semi-major axis ratio of known giant planet pairs 
that satisfy these criteria,
identified in the Exoplanet.org database
 as of May 2016.
We choose the giant planet orbits to be circular, but give each an
inclination of $5^{\circ}$, and an offset in the longitude of nodes of $\Delta \Omega = 231^{\circ}$.
This will ensure some oscillations in the inclinations and test how much inner inclinations can
be pumped. Using equations~(\ref{io}) and (\ref{outerf}) this model system implies a resonant location
at $a_2 \sim 0.3$AU.

\begin{figure}
\includegraphics[width=\columnwidth]{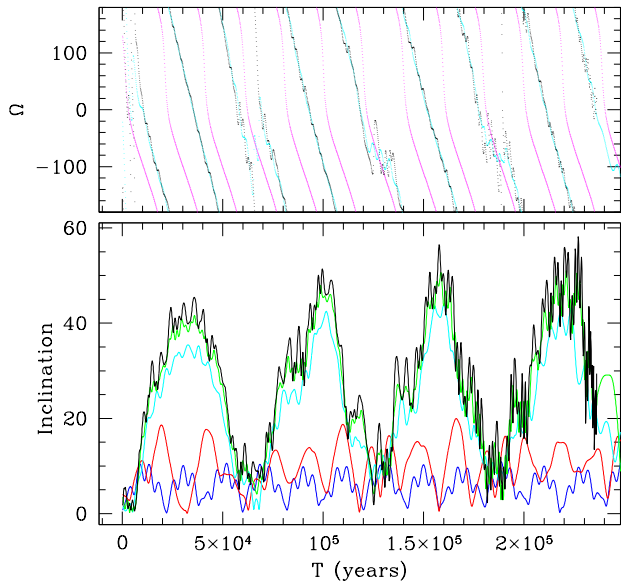}
\caption{The lower panel shows the oscillations in inclination for the same system
as shown in Figure~\ref{fig:ibounce}, but now subject to the inclination pumping of an external pair of
Jupiter mass planets. The colours are the same as before (the inner three planets are represented
by black, green and cyan respectively, and red indicates the fourth planet from the inside), with
a blue curve added to represent the outermost (fifth) of the non-Jovian planets. We see the inner
three planets still oscillate coherently under the inclination pumping, but to higher amplitudes
which eventually lead to eccentricity growth and crossing orbits (the inner pair collide at the
right edge of this figure). The upper panel shows the precession of the line of nodes for the innermost
and third planets, as well as for the innermost of the perturbing Jupiters (magenta curve). 
The similarity of the precession rates is responsible for the enhanced amplitude of the oscilations.
\label{fig:ibounce3}}
\end{figure}

The results are dramatic. Of the 50 systems simulated, only 4 retained all of their original planets.
The pumping of the inclinations is much stronger now, as expected, and most systems reach the
point where the inclinations do indeed get large enough to lead to eccentricity growth and
dynamical instability. As an illustration, Figure~\ref{fig:ibounce3} shows the same planetary system
as shown in Figure~\ref{fig:ibounce}, but now subject to the pumping of inclinations. The inner trio
of planets oscillate quite coherently as a unit, but eventually the eccentricity grows too and the
two inner planets cross orbits and collide. This spurs further instability and ultimately the three
inner planets coalesce into one and the outer two planets also collide, leaving a widely separated
two planet system.

The consequence of this inclination pumping and dynamical instability is that the surviving planets
are reduced in multiplicity (now we find 80\% of surviving systems have an
underlying multiplicity of 1--3 planets) and receive a larger dispersion in inclination and eccentricity. When
we observe the tranets, we find no more triple tranet or higher systems, and only a few double tranet
systems. The final $f_2/f_1=0.072\pm 0.001$, which is now quite a bit smaller than observations, offering the
possibility that we might combine this population with an undisturbed component to match the observations with a lower
overall occurrence of perturbers.
It also leaves a substantial inclination amongst the surviving planets. The median obliquity of the
surviving single tranets is $21^{\circ}$, while that of those found in multiple tranet systems
is $13^{\circ}$. This is now demonstrably larger than the inclination of the perturbers ($5^{\circ}$)-- indicative
of both the effects of resonant pumping and scattering in dynamical unstable systems. We also find
that $f(50|10)=0.73\pm 0.07$ for the single tranet systems, so the reduction in the overall multiplicity
of tranets yields a slight improvement in the period distribution for single tranets, but not enough to
get close to the observed value.

\subsection{Resonant Driving of Eccentricity}
\label{MultiPerte}

The level of dynamical instability observed here prompts the question whether the results are 
indistinguishable if we pump eccentricity or inclination. To test this, consider now Experiment~D1,
in which we retain the same giant planet pair, but now we make it almost coplanar (inclinations of 
$1^{\circ}$) but give each planet an eccentricity $e=0.2$ and $\Delta \bar{\omega} = 100^{\circ}$.
This will now pump eccentricity directly, and any dispersion in the inclinations will be the result
of planet scattering. The equivalent resonant location for Laplace-Lagrange oscillations in this case corresponds to $\sim 0.25$~AU.

Once again, we find a substantial amount of dynamical instability. In this case we find that only two
of our 50 systems retain all of their initial planets. Indeed, 13 of the systems lose {\em all}
their planets. This correlates with an increase in the number of collisions between planets
and the host star in this experiment. Over the 50 systems, we find 75 planet-planet collisions,
31 planet ejections, 1 collision with one of the Jupiters and 51 collisions between one planet
and the star. In the cases where the system loses all the planets, the last surviving planet finds
itself with a semi-major axis $\sim 0.2$~AU, which makes it vulnerable to eccentricity pumping
due to secular driving by the external Jupiters. This drives the eccentricity to large values and
eventual merger with the host star. There is reason to be cautious about taking such results
seriously as hybrid symplectic integrators such as {\tt Mercury} become unreliable with very
large eccentricities and the star-grazing periastron passages are not very well resolved even with our 
short timesteps. However, we have re-run the late stages of these simulations with a Bulirsch-Stoer
integrator and shorter timestep (see Appendix~\ref{RLO}) and verify that the stellar collisions are reproduced with
fidelity. Indeed, it is an interesting question whether these planets are truly lost, or whether they
undergo Roche-Lobe overflow and become captured in very short period orbits. This might provide
a mechanism for the production of the system of very short period orbits found
in the Kepler data (Rappaport et al., 2012, 2014; Sanchis-Ojeda et al. 2013, 2015;
Jackson et al. 2013).

Once again, we observe a dramatic reduction in the frequency of multiple tranet systems, although
we do retain some triple and quadruple systems in this case. The final ratios are
$f_2/f_1=0.171\pm 0.002$, $f_3/f_2=0.346\pm 0.022$ and $f_4/f_3=0.126\pm 0.008$. The value of $f(50|10)=1.86\pm 0.11$ for the single tranets
(1.22 for multi tranet systems), which is a dramatic improvement over previous simulations. Thus,
the resonant pumping of eccentricity appears to offer a way to reduce the excess of single tranets
at short periods. In this case, the change is clearly the result of a reduction in multiplicity,
as the median inclination of the single tranets is only 3.7$^{\circ}$. 

The change in the value of $f(50|10)$ relative to the other simulations is striking, and it is of
interest to understand the origins. It appears to be related to the increase in loss of planets
due to accretion by the star. The resonant location for Laplace-Lagrange oscillations due to this
perturber pair is at $\sim 0.25$~AU and planets near this location and not strongly tied to other
modes are thus excited to high eccentricities. Some of these are driven onto radial orbits which
eventually impact the star but others collide with planets at shorter orbital periods and merge
to provide a pile-up of planets around $\sim 0.1$AU. The increase in the frequency of planets in
this location results in the change in the statistics. The dramatic change in $f(50|10)$ is
partly artificial because the pile-up straddles the 10-day orbital period boundary, but the sense
of the shift is real.

We can examine the effect of resonant location by running simulations with different spacings 
for the outer planet pair. If we run the same Experiment but now with the second perturber shifted
outwards to 4.5~AU (Experiment~D2), we find similar multiplicity statistics as in D1, but now
$f(50|10)=0.67 \pm 0.05$. The outward shift of the outer perturber lowers the frequency of the
perturber secular oscillations, shifting the Laplace-Lagrange resonance inwards and reduces the
number of collisions/mergers at 0.1~AU and out. If we instead move the outer perturber inwards
to 3~AU while retaining the same eccentricities (Experiment~D3) then we find even greater instability.
Few double planet systems survive ($f_2/f_1=0.0034\pm 0.0001$) and the value of $f(50|10)=4.5\pm 0.4$. This is
because the eccentricity excitation is moved to larger scales, increasing the cross-section for
collision and resulting in a substantial population of final planets in the range $0.1-0.2$~AU.
Thus, we find that the best way to reduce the short period bias of single tranet systems is to
pump the eccentricity on larger orbital scales, by virtue of closer giant planet pairs.

\subsection{Sub Saturn Companions}

Experiments C1 and D1 suggest that it is indeed possible to reduce the tranet multiplicity substantially
with a sufficiently strong secular pumping. However, it seems to require multiple giant planets, and not every
observed planet system is observed to contain multiple planets. Can we achieve similar results with
smaller mass planets? We have already noted that a single perturber of mass $\sim 0.1 M_J$ yields
notably weaker results but perhaps if only the outer member of the pair were smaller it would change the
resonant frequency but the interior precession would still be significant. Thus, in experiments~C2, C3 and D4,
we repeat experiments C1 and D1, but with the outermost external perturber mass of $0.1 M_J$ (and all other parameters
held the same).

In experiment C2, the inclination pumping with a smaller outer planet yields multiplicity ratios
$f_2/f_1=0.32\pm 0.003$, $f_3/f_2=0.301\pm 0.003$, $f_4/f_3=0.212\pm 0.002$ and $f_{4+}/f_4=0.071\pm 0.005$.
The lower outer planet mass implies a smaller precession rate (see equation~\ref{SimHarm}) for the pair,
which, in turn, will shift the resonant location inwards (equation~\ref{io}). Nominally one might hope that
this improves the coupling to short period planets and increase $f(50/10)$, but the lower mass implies a 
much weaker driving, and the actual $f(50|10)=0.37\pm 0.03$ for singles, i.e. slightly worse than the unperturbed
case. 

Of course, reducing the mass of one of the giant planets also shifts the frequencies of the two planet
system and consequently the location of any resonances in the inner system. Perhaps a more faithful
comparison is to simultaneously move the outer planet inwards to a location such that the precession
rate (equation~\ref{SimHarm}) is preserved for the smaller perturber mass. For a $0.1 M_J$ planet, this
requires that it be moved in to 2.934~AU. The results (denoting this Experiment~C3) suggest a slightly
 stronger effect than in Experiment~C2, but still much weaker than Experiment~C1. In particular
$f_2/f_1=0.235\pm 0.002$, $f_3/f_2=0.223\pm 0.005$, $f_4/f_3=0.249\pm 0.006$ and $f_{4+}/f_4=0.047\pm 0.002$.
Furthermore $f(50|10)=0.68\pm 0.05$, so there is little improvement in the period distribution either.

In experiment D4, we repeat the eccentricity excitation of Experiment~D1, but with the 0.1$M_J$ outer perturber.
The eccentricity excitations are smaller but still lead to some level of dynamical
instability and loss of at least one planet in 34/50 systems. However, this is a much weaker effect than
in experiment~D1 (and only 1 system lost all its planets), and the resulting multiplicity ratios are $f_2/f_1=0.329\pm 0.009$,
 $f_3/f_2=0.289\pm 0.003$,
$f_4/f_3=0.269\pm 0.007$, $f_{4+}/f_4=0.085\pm 0.007$. In both cases the results are a little smaller than for the unperturbed population but
still higher than observed at low multiplicities, and together they 
suggest that we do indeed need multiple massive planets to substantially degrade the multiplicity
of a planetary system. This is because we require planets massive enough to induce substantial precession on small
scales, but they must also be driven with a similar frequency, which requires another planet of similar size.

We must also question whether such close giant planets are sufficiently ubiquitous to meet our requirements.
Experiments C1 and D1 suggest that giant planets at $\sim 1$AU are capable of substantially disrupting compact
planetary systems, but can we do it with giant planets on larger scales? Figure~\ref{fig:Prec} indicates that
the resonant excitation of the planetary systems was being driven on relatively small scales ($\sim 0.2$ AU)
and this is supported by the numerical results. However, planets on larger scales can still potentially
drive instability if they couple to planets on scales $\sim 1$~AU. Thus, let us now consider experiments C4 
and D5, in which we repeat experiments C1 and D1 with more distant giant planet pairs.

\subsection{More Distant Planets}
\label{MoDis}

\begin{figure}
\includegraphics[width=\columnwidth]{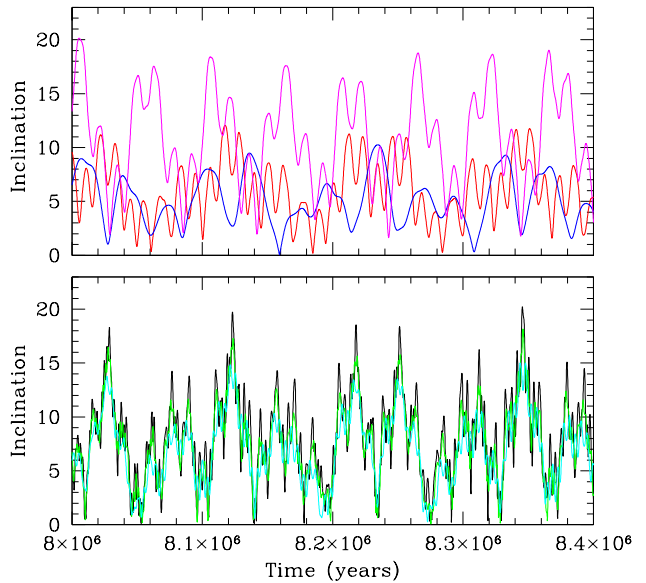}
\caption{The lower panel shows the oscillations in inclination for the three
inner planets (black, green and cyan curves). The upper panel shows the inclination oscillations
for the outer three planets of the system, with red and blue indicating the fourth and fifth
planets as before, and magenta now indicating the sixth planet, at 1.099~AU. It is this
last planet that is closest to resonance with the outer giant planets and experiences the
largest inclination pumping. 
\label{fig:ibounce4}}
\end{figure}

In experiment C4, we place two $1 M_J$ mass planets on circular orbits at 3 and 5.6~AU, with an inclination
of 5$^{\circ}$ and $\Delta \Omega = 100^{\circ}$. Because of the larger distance of the giant planets, we are now
able to use the full complement of originally simulated planets, including those with semi-major axis
$\sim $1~AU.
The resulting inclination excitations in this case are indeed weaker, but
not negligible, and 13/50 systems cross the line into dynamical instability to the point of losing at
least one planet. Figure~\ref{fig:ibounce4} shows the behaviour of the same system as shown before
in Figures~\ref{fig:ibounce} and \ref{fig:ibounce3}, but now with the more distant perturber. This system
now also includes a sixth planet at 1.099~AU. Once again, the inclination of the three innermost planets exhibit
substantial correlation but do not grow as large as in Figure~\ref{fig:ibounce3}. 

 The resulting multiplicity ratios averaged over the 50 systems  are
$f_2/f_1=0.255\pm 0.003$, $f_3/f_2=0.231\pm 0.002$, $f_4/f_3=0.260\pm 0.005$ and $f_{4+}/f_4=0.317\pm 0.012$. Thus, it does reduce the
tranet multiplicity, but not as much as when the planets are closer. Also $f(50/10)=0.56\pm 0.04$ for the single tranet systems, so the
results are still well skewed to short periods.

\begin{figure}
\includegraphics[width=\columnwidth]{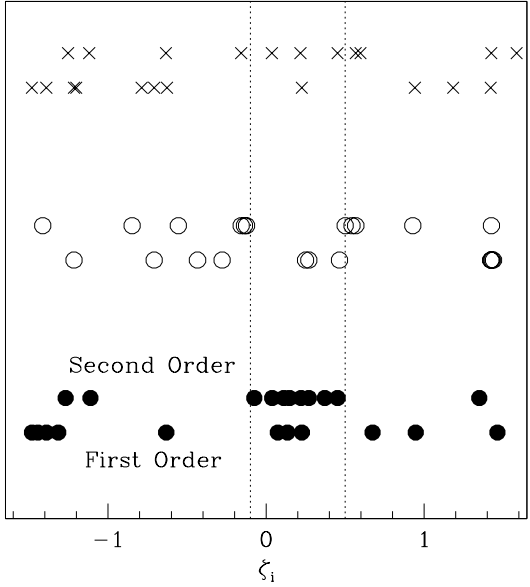}
\caption{The solid points at the bottom of the plot show the values of $\zeta_1$ (lower line) and $\zeta_2$ for the
outer pair in the 11 systems where no planet was lost. The open circles and crosses show the same quantities for two other samples
of 11, drawn randomly from the initial conditions in systems where a planet was lost. The robust systems seem to show a concentration
of $-0.1 < \zeta_2 < 0.5$.
\label{fig:Stats}}
\end{figure}

In experiment~D5, we retain the masses and seperations of C4, but 
 we pump the eccentricities instead of the inclinations, by giving the giant planets $e=0.2$, $\Delta \bar{\omega}=100^{\circ}$
and inclination of $1^{\circ}$ ( as in experiment~D1). As expected, we find a greater level of dynamical instability. In 39/50 systems we see
the loss of at least one planet, often several. We see a total of 76 planet collisions, 35 collisions with the star, and 7 ejected planets
over the 50 simulations. We also see one case of a planet impacting one of the Jupiters and one case where the interactions destabilise
the outer giant planet pair, leading to the ejection of the outer giant planet.
On the other hand, in 11 cases, we see no loss of planets at all. One pattern that emerges is that the resilient systems tend to have
 planet pairs that are close to second order mean motion resonances on scales $\sim 1$AU. To quantify this, Figure~\ref{fig:Stats} shows the distribution
of the statistics $\zeta_1$ and $\zeta_2$, as defined by Lissauer et al. (2011), that measures the proximity to first order and second
order resonances respectively, calculated for the outermost pair of each inner system (i.e. not counting the perturbers). 
The solid points show the values for the 11 such pairs in the systems in which no planet was lost.
The $\zeta_1$ values are spread almost uniformly but $\zeta_2$ shows 8/11 with values $-0.1 < \zeta_2 < 0.5$. If we draw 11 samples
randomly from a distribution function fit to the observed Kepler distribution (discussed in \S~\ref{Prats}) we find that this happens
with a frequency of only $5 \times 10^{-5}$. However, this distribution function does not take into account possible underlying
trends with semi-major axis and systematic effects, so assigning a statistical significance is questionable. Another test is
to draw additional samples of 11 from the initial conditions of those systems that did lose at least one planet. Two such samples
are shown as crosses and open circles in Figure~\ref{fig:Stats}. In these cases we find 1/11 and 3/11 with  $-0.1 < \zeta_2 < 0.5$.
Such samples should occur with frequencies of 0.29 and 0.12 if we use the distribution function. Thus, it does appear as though
the surviving systems have an excess of outer pairs near a second order resonance (most of these are the 3:1, although there is
one near 7:5). 

 In the case of first order resonances, there is a well defined potential mechanism for this protection.
 The correction to the secular frequencies that results from such
interactions (Malhotra et al. 1989; Christou \& Murray 1997) raises the precession rates and  makes the system more robust to the external perturbations
by shifting the system away from resonance. A similar mechanism appears to be operating for the second order resonances,
although the formalism does not translate directly to this case.

In terms of the statistics, $f_2/f_1=0.240\pm 0.005$, $f_3/f_2=0.319\pm 0.003$, $f_4/f_3=0.236 \pm 0.004$
and $f_{4+}/f_4=0.32 \pm 0.01$.
We also see $f(50/10)=0.75\pm 0.04$ for singles, which is closer to the expectations.

\begin{figure}
\includegraphics[width=\columnwidth]{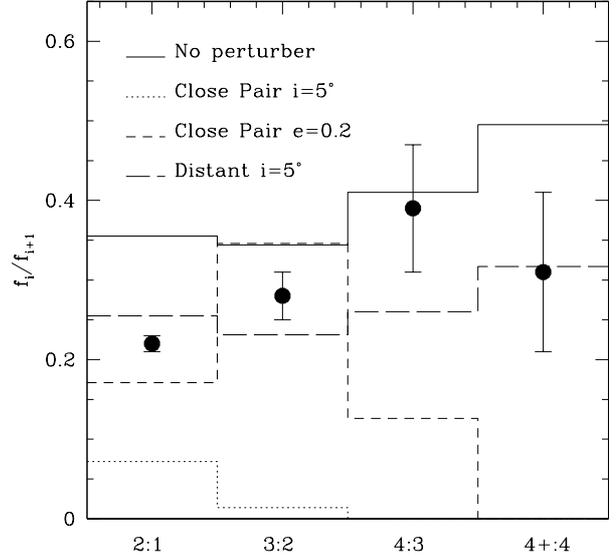}
\caption{
The solid points shows the multiplicity ratios of the cumulative Kepler candidate
sample, as of July 2016, with Poisson
error bars, when considering only systems in the period range 5--260~days (the most
accurate comparison to our simulations). 
 The solid histogram is the expectation from our model simulations without
any external perturber, when
observed from random orientations. The dotted histogram shows the result of inclination pumping by a
pair of $1 M_J$ planets (Experiment C1), while the short dashed histogram show the effects of eccentricity
pumping by the same pair (Experiment D1). Both of these have strong effects on the tranet multiplicity.
The effects of inclination pumping by a more distant pair (Experiment C4 -- long dashed histogram) are weaker
but do reproduce the flatness of the observational trend.
\label{fig:Rat4}}
\end{figure}

Figure~\ref{fig:Rat4} shows a summary of the effects of pairs of perturbers on the tranet multiplicities.
We see that pairs with the inner member interior to 2~AU can have significant effects, and can
bring the single tranet fraction well below the observed limits. More distant pairs do reduce the
multiplicity somewhat, but not enough to match the observations. 

\subsection{Synthesis}

Up to now we've performed simulations of simple example cases, in which we look at the influence of different sets of
perturbers, with fixed parameters within each experiment.
What does the result for a plausible range of external parameter choices look like? The eccentricity
distribution can be characterised from the results of radial velocity observations, but the distribution
of inclinations is poorly known. We thus need a theoretical model that links both the eccentricity
and inclination distributions.

\subsubsection{Single Planet Perturbers}
\label{JT}

Juric \& Tremaine (2008)
examined the hypothesis that the observed distribution of giant planet eccentricities was the result
of planet-planet scattering and report the resulting distributions of both eccentricity and inclination.
 This gives us
a set of perturbers whose eccentricities are chosen to match the observed distribution and which also
possess a plausible range of inclinations, at least under the hypothesis that both are the result of
planetary scattering.

Thus, 
we have also performed Experiment~E1, in which we place a single $1 M_J$ planet at 2~AU, and draw our
inclinations and eccentricities from the final results reported in Juric and Tremaine. 
 We have moved
the perturber out to 2~AU because the eccentricities are larger than $e=0.2$ in many cases, and may
remove many planets by direct scattering in that event, if started from 1.25~AU. 
We choose both eccentricities and inclination from a distribution function of
the form $p(x) \propto x \, exp(-1/2 (x/x_0)^2)$, where $e_0=0.3$ and $i_0=7^{\circ}$\footnote{We
also renormalise the eccentricity distribution to account for the limit that $e<1$, but this
is an effect $< 1\%$}.

As might be expected from a broader distribution, we see a full range of behaviours. Indeed
14/50 systems retained all their planets, while 3/50 lost all their planets. The final statistics
are $f_2/f_1=0.164\pm 0.003$, $f_3/f_2=0.143\pm 0.004$ and $f_4/f_3=0.039\pm 0.001$. Thus, there is
a substantial effect. The ratio of $f(50|10)=0.87 \pm 0.05$ for surviving single tranet 
systems. The obliquities of the surviving systems are not substantially larger than for prior
experiments either, suggesting that the reduction is mostly the result of pumping eccentricity
to the point of dynamical instability.
In all, these 50 systems lose 63 planets to planet-planet collisions, 37 to collisions with the star,
5 to ejection and 1 to a collision with the perturber. 

These results suggest that introducing a giant planet perturber with the sort of amplitudes
suggested by radial velocity observations can reduce the tranet multiplicities substantially, but
it would require a large fraction of planetary systems to be subject to this kind of perturbation,
much as in the case of Experiment~A1. 

\subsubsection{Planet Pair Perturbers}

As we have seen, giant planet pairs have even greater effects by virtue of their ability to resonantly
pump modes in the compact planetary system. To generate an planet pair equivalent to the single
planetary systems of E1, we proceed as follows. We shall assume both planets have
mass $1 M_J$, and that the inner planet is again located at 2~AU. The separation $\alpha = a_1/a_2$ is drawn uniformly from between 0.2 and 0.6, which
encompasses the majority of known pairs whose inner member lies at semi-major axes $> 1$AU. We then
select eccentricities and inclinations from the same distributions as in \S~\ref{JT}, and select
the longitudes of periastron and ascending node randomly. For each pair we then integrate them
with Mercury for $10^5$ years and discard any systems whose inner semi-major axis has deviated by more
than 1\%. In this fashion we retain only those whose evolution is secular in nature, and which are
not evolving by planetary scattering. The surviving pairs are more skewed to small $\alpha$ than the
original uniform distribution (median value $\alpha=0.315$) because closer pairs are more likely to
be drawn with eccentricities such that orbits cross and thereby violate the secular assumption.
The initial eccentricity distribution for the inner planet is also skewed to smaller values than
in Experiment~E1, with fewer eccentricities $> 0.4$.

We then combine the resulting perturbing systems with our inner planet systems and integrate,
making this Experiment~F1. As in the previous cases of pairs of perturbers, the results are stronger
than in Experiment~E1. 
In this case 18/50 of the systems lose all their planets, and most (27) of the systems with a surviving
interior planet are singles. The ratio of $f_2/f_1=0.028 \pm 0.002$, with $f_3/f_2=0.129 \pm 0.001$. It is also notable
that there are as many planetary collisions with the star as there are planet-planet collisions. 
This certainly provides a distinct population of single tranets, but unfortunately the extreme dynamical
instability yields $f(50|10)=0.39 \pm 0.03$, i.e. the survivors are even more biased to short periods than
the original systems. This is because the high eccentricities bias the systems to large $\alpha$ when we
impose conditions of dynamical stability, and this favours low $f(50|10)$ (see \S~\ref{MultiPerte}).

This configuration is perhaps the most dynamically excited possible, given the constraints of long-term
stability (at least for the giant planet pair). We also test a less extreme version, Experiment~F2, in which
we repeat this experiment but draw the eccentricities and inclinations for the giant planets from distributions
with dispersions half that of those in Experiment~F1, although the separations are drawn from the same uniform
distribution, and the inner Jupiter is still located at 2~AU. We also still test each planet pair for
dynamical stability before combining it with an inner system.

Experiment~F2 yields $f_2/f_1=0.085\pm 0.003$, $f_3/f_2=0.152\pm 0.008$ and $f_4/f_3=0.021\pm 0.001$. The
period distribution for singles yields $f(50|10)=0.67 \pm 0.05$, so definitely an improvement over Experiment~F1,
but only a modest improvement over the unperturbed population, and not enough to match the observations.

How far out can a perturbing system be while still affecting the inner planets? As we observed in \S~\ref{IncPert},
the effects of an distant inclined perturber is limited because the entire inner system tends to precess as a whole.
On the other hand, we have found that eccentricity pumping is an important component although the results of
Experiment D5 suggest a limited effect. Nevertheless, Experiment~F1 shows that exploiting the full range of
eccentricities and inclinations motivated by observations strongly affects interior planets. Thus, to assess the
effectiveness of distant, dynamically excited populations, we consider Experiment~F3, in which we postulate a pair
of more massive $3 M_J$ planets, with the inner member at 5~AU. This combination is chosen on the basis of the
considerations shown in Figure~\ref{fig:Prec}, such that the combination contributes competitively to the precession
on scales inside 1~AU. The semi-major axis of the outer member is then chosen by sampling a uniform distribution
in semi-major axis ratios for the pair, and the eccentricities and inclinations are chosen from the undiluted distributions
from Juric \& Tremaine. As before, we reject those combinations which yield dynamically unstable configurations.

The results suggest that massive, distant companions can still exert a substantial influence, with
$f_2/f_1=0.121\pm 0.004$, $f_3/f_2=0.228\pm 0.004$, $f_4/f_3=0.127\pm 0.003$ and $f_{4+}/f_4=0.083\pm 0.006$.
This is indeed weaker than the corresponding closer case in Experiment~F1, but still sufficient to lower the
single tranet frequency below the observed value. Thus, we conclude that, once we account for the full range
of observed eccentricities (and equivalent inferred inclinations), giant planet perturbers as distant as
5~AU can have a substantial influence on the compact planetary systems.

\section{Two Populations}
\label{TwoPop}

Up to this point, we have calculate the multiplicities for a variety of single populations,
finding that some classes of perturber can reduce the multiplicities to observed values
and maybe even lower them below the observed value. A more likely prospect is that the
observed population is composed of multiple subpopulations. In particular, we want to consider
here the case where some fraction of the population is composed of unperturbed, high multiplicity
systems and the rest is drawn from one of the perturbed populations discussed above.

If we adopt a model in which the compact planetary systems all form by in situ assembly, but some fraction
$f_J$ is accompanied by external perturbations, then we can ask what combination is needed to match the
multiplicity ratios? Let us assume the undisturbed model produces N planetary systems. The apportioning into different
tranet multiplicities is such that
\begin{eqnarray}
N & = & N_1 + N_2 + N_3 + N_4 + N_{+} \nonumber \\
  & = & N_1 \left( 1 + \frac{N_2}{N_1} + \frac{N_3}{N_1} + \frac{N_4}{N_1} + \frac{N_{4+}}{N_1} \right) \nonumber \\
  & = & N_1 \left( 1 + \frac{f_2}{f_1} + \frac{f_3}{f_2} \frac{f_2}{f_1} + \frac{f_4}{f_3} \frac{f_3}{f_2} \frac{f_2}{f_1} +
 \frac{f_{4+}}{f_4} \frac{f_4}{f_3} \frac{f_3}{f_2} \frac{f_2}{f_1} \right) \nonumber \\
  & = & N_1 \left( 1 +  0.36 + 0.34 \times 0.36 + 0.40 \times 0.34 \times 0.36 + \right. \nonumber \\
 &&  \left. 0.50 \times 0.40 \times 0.34 \times 0.36 \right) \nonumber \\
  & = & 1.556 N_1
\end{eqnarray}
where $N_1$ is the number of observed single tranet systems, $N_2$ is the number of double tranet systems, etc. The equivalent ratio
between the total number of transitting systems to those with only single tranets is 1.552 (or, equivalently, an average of 1.546 tranets 
peri transitting planet
system).

One can calculate a similar relationship for the systems that suffer perturbations. Let us consider Experiment~F2 as an example
of a perturbing population. In this case, $N=1.098 N_1$, because there are far fewer multiple transit systems in the perturbed sample.
 If we combine this with
the unperturbed systems in the proportions required
to match our estimated observed value of $f_2/f_1=0.22 \pm 0.01$, then we require
\begin{equation}
\left(\frac{N_2}{N_1}\right)_{obs} = \left( \frac{f_2}{f_1} + \left(\frac{f_2}{f_1}\right)' \frac{N'_1}{N_1}\right) \left( 1 + \frac{N'_1}{N_1} \right)^{-1}
\end{equation}
where $N_1$ is the number of singles from the undisturbed population and $N'_1$ is the number of singles from the disturbed population (Experiment~F2 in this case).
To match the observations requires $N'_1/N_1 = 1.0 \pm 0.08$, which corresponds to a ratio between the total number of
perturbed and unperturbed systems $N'/N = 0.71 \pm 0.06$. Thus, a mixed population in which $f_J \sim 40\%$ of the underlying systems
are drawn from Experiment~F2 (and the rest are unperturbed) will match the observations.
 If we adopt $N'=2/3 N$ as a baseline model, we get $f_2/f_1=0.224 \pm 0.002$,
$f_3/f_2=0.309\pm 0.003$, $f_4/f_3=0.375\pm 0.004$ and $f_{4+}/f_4=0.493 \pm 0.004$.

We can calculate similar mixtures for any case in which the perturbed population yields a value $f_2/f_1$ smaller than
the observed value. In some cases, such as Experiments~A1, D1 \& E1, matching the observations require that most of the observed
planetary systems correspond to the perturbed case. In others, such as A4, D3 or F2, we can match the observed single tranet
excess with populations that feature non-negligible fractions of unperturbed systems.
 Let us construct three illustrative  examples. First, consider
the case in which we select 30\% of the systems from Experiment~D3 and the rest from the unperturbed sample. This represents
an attempt to match the tranet frequencies while maximising the value of $f(50/10)$.
 The results of Experiment~D3 overwhelmingly produce 
single planets, and these systems produce 40\% of the single tranet systems in this combination, leading to $f_2/f_1=0.203 \pm 0.004$.
This combination also produces $f(50|10)=1.13 \pm 0.07$, which is closer to the observed value that any other combinations that
match $f_2/f_1$. 
A second example model is the case in which we
select 30\% of systems from experiment A4 and the rest from the unperturbed population. This provides $f_2/f_1=0.218\pm 0.002$. The difference in this case is that the perturbers
are small enough to escape detection (those of Experiment~C1 or F2 are probably not) through radial velocities. The value of $f(50|10)$ is not as good a match, but
this combination does also provide a potential signature in that the obliquity distribution of the single tranets is much wider than that of the
multi-tranet systems. The same is not true for the model with a D3 component, indicating how we might eventually tease out the different contributions.

\begin{figure}
\includegraphics[width=\columnwidth]{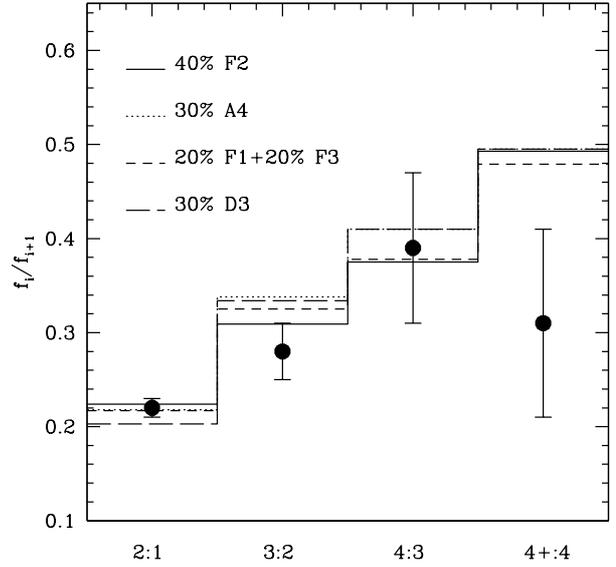}
\caption{The points shows the multiplicity ratios of the Kepler candidates
when we restrict the sample to orbital periods 4--260 days. The short dashed histogram shows
the multiplicity ratio that emerges from a population drawn in equal proportion from the
unperturbed population and from Experiment~C1. Similarly, the dotted histogram represents
a population drawn 40\% from 
 Experiment~C4.
 The long dashed histogram is drawn 90\% from Experiment D1, which is the case that matches
the observed period distribution statistic $f(50|10)$.
Finally, the solid histogram is constructed from a population that is 50\% from Experiment~F2
and 50\% from unperturbed systems. 
\label{fig:Rat5}}
\end{figure}

Finally, it is also possible to match up combinations from different experiments, to mimic broader
distributions that are more likely to represent reality.
 Let us consider a combination of 2/3 from Experiment~F1 and 1/3 from
Experiment~F3. This can represent a perturbing population with giant planets at a range of
distances. In this case, $f_2/f_1=0.065 \pm 0.002$, which is still well below the observations.
However, we can now combine this mixed population with the unperturbed systems to match the
observations. In this case, we once again consider $f_J =  33\%$ (split 2:1 between F1 and F3), yielding
$f_2/f_1=0.240 \pm 0.002$, $f_3/f_2=0.328 \pm 0.003$, $f_4/f_3=0.373 \pm 0.004$ and
$f_{4+}/f_4=0.487 \pm 0.005$. In this case, $f(50|10)$ is still too low, but there is still
a potentially measurable difference in obliquities between singles and multiples.
 Figure~\ref{fig:Rat5} compares these combined models to the observations. They are all chosen
to reproduce the ratio $f_2/f_1$, and then provide a range of values for higher multiplicities.

Thus, we find that it is possible to at least get close to the observations by combining the results
of different experiments. However, different choices imply different results for other observables,
which may offer additional pathways to discerning which effect actually dominates the observed systems.

\section{Longer Term Effects}
\label{LongView}

We have thus far considered only the effects of secular pumping on timescales of 10~Myr, and this certainly
proves to be enough time to substantially modify many of the observed systems. However, this does not
exclude the possibility of further evolution due to the accumulation of weaker nonlinear effects on Gyr timescales.
Thus, we must consider whether the systems modelled here will change their character significantly before
we observe them, due to the growth of further, longer-term dynamical instabilities.

On the other hand, many of these planetary systems still retain planets on small orbital scales, where
tidal effects can serve to damp eccentricities. Indeed, the combination of tidal dissipation and secular
coupling can serve to stabilise many systems against further dynamical evolution (HM15).

\subsection{Tidal Evolution}
\label{TidEv}

As noted by several authors (Greenberg \& van Laerhoven 2011; Laskar, Boue' \& Correia 2012), the combination of secular coupling and tidal evolution
can result in the damping of eccentricity of several planets, not just the innermost. HM15
investigated this effect in the context of the unperturbed systems discussed here and concluded that tides
which circularise individual planets out to periods $\sim 10$~days could
potentially circularise  orbits in secularly coupled systems out to periods $\sim 100$~days.
The increased dynamical excitation in the systems studied here  leads to higher eccentricities
and thus can potentially drive stronger evolution. A full examination of the consequences would require
estimates of the strength of tidal dissipation and is beyond the scope of this investigation. However, HM15
described the basic character of the unperturbed systems and their evolution under the effects of tides.
We can assess the broad effects of tidal disspation by applying the following conclusions of HM15 to our results. 

We assume that, if the innermost surviving planet in the system has a periastron $< 0.07$AU, it has the
potential for tidal circularisation. This is estimated on the basis of the formalism outlined in 
Hansen (2010), assuming a planet of $5 M_{\oplus}$ and $ 2 R_{\oplus}$ with an eccentricity of 0.5 and
a strength of tidal dissipation corresponding to Q=1000 (which is a relatively weak dissipation compared
to Earth but comparable to that of  Mars, which is probably a better analogue for desiccated rocky planets like those 
on short period orbits). In that event, we circularise all surviving
planets with orbital periods $<$100~days, and move the innermost planet inwards to conserve the total
system angular momentum.

In cases such as Experiments  A3, A4 or F1, the dynamical evolution is so extensive that most surviving
systems are either single planets or widely separated multiples, so that secular interactions have
little influence on the tidal evolution. The largest effects are seen in cases where the eccentricities
are excited to significant levels, but low enough that the system retains some level of multiplicity.
In this case, secular interactions can continually pump the eccentricity of the inner, tidally evolving
planet, thereby circularising all planets with substantial secular interactions with the inner planet
(as described in HM15).
 In the case of Experiment~D1 for example, 20/50 systems experience some level of inward migration due
to tidal evolution. In some cases, this is a relatively minor shift, but, in other cases, the inner planet can
migrate well inside the nominal inner edge of the initial conditions. The most dramatic case involves a
system in which three planets survive in addition to the two perturbing Jupiters. The innermost is a 2.8$M_{\oplus}$
planet in a 5.4~day orbit, while another is 7.2$M_{\oplus}$ planet on a 28.3~day orbit. This second planet has an
eccentricity of e=0.504 at the end of the simulation and consequently exhibits a substantial angular momentum
deficit, which the tidal migration of the smaller planet must fill (the third planet has an orbital period of 154.6~days
and is essentially decoupled). In order to completely circularise the second body, the innermost must migrate
down to an orbital period of 7.1~hours! Such an evolutionary pathways offers another route to the very short
period planets observed by Kepler (Rappaport et al., 2012, 2014; Sanchis-Ojeda et al. 2013, 2015;
Jackson et al. 2013).

Indeed, this and other systems with substantial excitation provide a mechanism to populate orbits with orbital periods
too short to be populated by in situ accretion -- an objection to this model raised in the past by some authors,
such as Swift et al. (2013). To illustrate, we can calculate $f(4|10)$, a statistic similar to $f(50|10)$ but consisting of
the relative fraction of all single tranets observed with periods $< 4$~days compared to those with orbital periods between
4--10 days. If we apply the above procedure to all the systems in Experiment~D1 and repeat the estimates of transit
probabilities, we find that $f(50|10)$ is reduced to $1.03\pm 0.07$ but that $f(4|10)=1.03 \pm 0.08$, substantially higher
than the value of $0.09 \pm 0.02$ before the correction for tidal evolution is applied. The corresponding number for the
unperturbed population is $f(4|10)=0.06 \pm 0.01$ and the observed value (if we relax our period cut but retain the
stellar and magnitude limits) is $f(4/10)=0.80 \pm 0.05$. Thus, tidal evolution can, in principle, populate shorter
periods at the needed rate.

The experiments drawn from the Juric \& Tremaine distributions also show a healthy population of
short period orbits. Experiment~F1 shows  $f(4|10) = 0.58 \pm 0.02$ even before the inclusion of
tidal evolution because of the large level of dynamical excitation in this experiment. The equivalent
values for Experiments F2 and F3 are 0 and $0.042 \pm 0.008$ respectively. However, after the inclusion
of tidal evolution, all three experiments show healthy populations of tranets with P$<$4~days. In
the case of Experiment F1, $f(4|10)=0.90 \pm 0.06$, while it is $0.27\pm 0.03$ and $0.52 \pm 0.04$
for Experiments F2 and F3. Thus, the level of dynamical excitation observed in these experiments is sufficient
to generate numbers of short period planets comparable to that observed, if we include the effects of
tidal evolution.

\subsection{Longer Term Chaos}

The calculations described here focus on the dynamical instabilities that result if the external secular pumping
were sufficiently strong to lead to orbit crossing within 10~Myr. It is still an open question as to whether longer
term evolution may lead some additional systems to reach instability by virtue of smaller, nonlinear effects that
may slowly lead to eccentricity growth and orbit crossing. Such `secular chaos' may operate in a variety of planetary
systems (e.g. Lithwick \& Wu 2014).

We can estimate the sensitivity of our systems to long-term chaos by calculating the total angular
momentum deficit for the system (Laskar 1997), as
\begin{equation}
\Lambda = \sum_k m_k \sqrt{a_k} \left( 1 - \sqrt{1-e_k^2} \cos i_k \right) \label{amdeq}
\end{equation}
where the sum is over all the planets in the system. As discussed by Laskar, this quantity is 
approximately conserved in a global sense during the system evolution. If nonlinear effects allow
mixing of this deficit amongst the planets, one could imagine a situation where, in principle, all
but one of planets in the system instantaneously occupy circular, coplanar orbits, resulting in the
concentration of $\Lambda$ in the orbit of a single planet, yielding a maximal eccentricity for
that planet. If this maximum eccentricity allows for crossing orbits then planet-planet scattering
can break the secular assumption and lead to further dynamical instability. If orbit crossing is
not achieved even under such extreme circumstances, the system is probably stable in the long term.

If we calculate the possibility of orbit crossing in this manner for the 50 unperturbed systems
used here, we find that 23/50 have the possibility of instability, given the parameters that
emerge from the dynamical evolution.
However, as discussed in HM15, the
value of $\Lambda$ is probably diminished substantially by the combined effects of tidal evolution
and secular interactions, which will have the tendency to circularise the orbits of all planets
that are secularly coupled to the inner planet, rendering such systems stable. If we set the
eccentricities (but not inclinations) in the calculation of $\Lambda$ to be zero from all planets with
 $a<0.4$AU (essentially the prescription discussed in \S~\ref{TidEv}), then we are left with only 3/50
of the original unperturbed systems with even a small likelihood of longer term orbit crossing.

For the perturbed systems, we do not include  the giant planet perturbers in the calculation
of equation~(\ref{amdeq}). To do so would automatically render a system potentially unstable, because the large mass of
the perturbers will overwhelm any contributions from the interior planets. However, the limited expected
feedback from the small planets on the large means that we do not realistically expect to feed all
of the angular momentum deficit of the large planets into the interior system. This is similar to the
rationale of Laskar (1997) for separating the inner and outer Solar System planets.

Can longer-term dynamical instability lower the multiplicity for those experiments which gave too
large a value of $f_2/f_1$? For example, if
 we calculate the potential for stability in the results of Experiment~A1, the increased excitation
of eccentricity and inclination allow for the possibility of long-term instability in 40/50 of the
modelled systems, although this decreases to 13/50 if we estimate the effects of tides as above.
In the equivalent case of eccentricity forcing, Experiment~B1, 
 the number of potentially unstable systems is only 12/50 and none satisfy
the criteria if we take tidal damping into account. The tendency to greater stability is, in part,
due to lower surviving multiplicity, but also because of the lower inclinations in this case.
The results of all the experiments show the same trends -- those with moderate
excitation produce  many systems with global $\Lambda$
that can, in principle, lead to instability on Gyr timescales, but most are stabilised by the effects
of tidal damping because the secular coupling extends the reach of circularisation out to periods
$\sim 100$~days. Furthermore, the experiments with the greatest level of dynamical excitation produce
surviving systems that are sufficiently sparse that they are unlikely to be unstable. For instance,
of the 45 surviving planetary systems produced by Experiment~F2 (5 lost all their original planets), none have a large enough $\Lambda$ value,
even before tides are included.

Thus we conclude that longer-term chaotic evolution of these model systems are not likely to
substantially change the conclusions we draw. While some
potential for further dynamical evolution certainly remains (although the above estimate somewhat
overstates the likelihood) in many of the modelled systems, it is lower in those experiments that
do a good job of matching the observations. While further evolution could potentially improve the
agreement with observation in some experiments (like A1 or C1), it must also be noted that the effects
of tides operate in precisely the region (orbital periods $< 50$~days) that is most important for
comparing to observations and so reduce the likelihood of substantial evolution. Thus, we will assume
that the estimates based on our simulations hold even after $\sim 1$Gyr of further secular evolution.

\section{Closer Comparisons to the Data}
\label{DatComp}

We have seen in the above sections that it is possible, at least in principle, to reduce the
tranet multiplicity in our model systems to the point that it can match the observed distribution.
However, it is now necessary to ask whether the kinds of perturbations required are reasonable
or can be tested by other observations.

Table~\ref{tab:TranTab} collects the data on the various comparisons between the model and the
data.
\begin{table*}
\centering
\begin{minipage}{180mm}
\caption{{\bf Frequency of Tranet Multiplicity}. The second line shows the multiplicity drawn from
the subset of Kepler planets that meet our period cuts between 4 and 260~days. The parameters
of each experiment are described in the text. The number $<N>$ is the average number of tranets
per star for each model. We also show the middle 68\% range of inclinations for single tranets ($I_s$)
and tranets in multiple systems ($I_m$) as well as the equivalent ranges of eccentricity 
($e_s$) and ($e_m$). In these latter cases, we consider only planets with orbital periods $> 100$~days,
as shorter periods are likely to be tidally circularised.
 \label{tab:TranTab}}
\begin{tabular}{@{}lcccccccccc@{}}
\hline
Name & $f_2/f_1$ & $f_3/f_2$ & $f_4/f_3$ & $f_{>4}/f_4$ & $<N>$ & $f(50|10)$ & $I_s (^{\circ})$ & $I_m (^{\circ})$ & $e_s$ & $e_m$  \\
\hline
Cumulative & 0.22(1) & 0.28(3) & 0.39(8) & 0.31(1)& 1.34  & 1.42(8) \\
DR24 & 0.21(1) & 0.27(4) & 0.38(8) & 0.35(12) & 1.32 & 1.43(8) \\
\hline
Unperturbed & 0.355(3) & 0.344(3) & 0.410(4) & 0.495(4) & 1.55 & 0.42(3) & 7(5) & 3(3) & 0.09(5) & 0.06(5) \\
Experiment~A1 & 0.184(2) & 0.114(2) & 0.063(2) & 0.417(9) & 1.19 & 0.56(4) & 13(9) & 10(8) & 0.05(2) & 0.05(2) \\
Experiment~A2 & 0.258(3) & 0.168(3) & 0.093(3) & 0.056(3) & 1.27 & 0.62(4) & 14(8) & 12(6) & 0.07(6) & 0.05(4) \\
Experiment~A3 & 0.080(1) & 0.035(1) & $\cdots$ & $\cdots$ & 1.08 & 0.40(3) & 29(14) & $\cdots$ & 0.12(8) & $\cdots$\\
Experiment~A4 & 0.0100(2) & $\cdots$ & $\cdots$ & $\cdots$ & 1.01 & 0.96(7) & 51(25) & $\cdots$ & 0.33(18) & $\cdots$ \\
Experiment~B1 & 0.288(2) & 0.271(8) & 0.278(6) & 0.129(4) & 1.37 & 0.62(4) & 5(4) & 3(3) & 0.14(8) & 0.12(9) \\
Experiment~B2 & 0.289(1) & 0.354(2) & 0.420(7) & 0.118(2) & 1.45   & 0.46(4) &  8(7) & 2(2) & 0.12(6) & 0.11(8) \\
Experiment~B3 & 0.300(6) & 0.197(5) & 0.163(6) & 0.156(8) & 1.33   & 0.66(4) & 7(6) & 2(2) & 0.16(9) & 0.05(4)\\
Experiment~C1 & 0.072(1) & 0.014(1) & $\cdots$ & $\cdots$ & 1.07 & 0.61(5)  & 21(11) & $\cdots$ & 0.14(11) & $\cdots$\\
Experiment~C2 & 0.324(3) & 0.301(3) & 0.212(2) & 0.071(5) & 1.41 & 0.37(3) & 9(7) & 4(1) & 0.08(6) & 0.05(4) \\
Experiment~C3 & 0.235(2) & 0.223(5) & 0.249(6) & 0.047(2) & 1.30 & 0.52(4) & 12(7) & 9(6) & 0.06(5) & 0.04(3) \\
Experiment~C4 & 0.255(3) & 0.231(2) & 0.260(5) & 0.317(12) & 1.33  & 0.34(3) & 9(6) & 5(4) & 0.09(8) & 0.05(4) \\
Experiment~D1 & 0.171(2) & 0.346(22) & 0.126(8) & $\cdots$ & 1.25 & 1.86(11) & 6(5) & 1.3(13) & 0.18(12) & 0.18(9) \\
Experiment~D2 & 0.174(4) & 0.217(2) & 0.092(3) & $\cdots$ & 1.21 & 0.67(5) & 7(6) & 4(3) & 0.20(9) & 0.12(1)  \\
Experiment~D3 & 0.00034(1)  & $\cdots$ & $\cdots$ & $\cdots$ & 1.01 & 4.5(4) & 7(6) & $\cdots$ & 0.22(9) & $\cdots$ \\
Experiment~D4 & 0.329(9) & 0.289(3) & 0.269(7) & 0.085(7) & 1.42 & 0.44(3) & 6(5) & 3(3) & 0.15(8) & 0.12(9) \\
Experiment~D5 & 0.240(5) & 0.319(3) & 0.236(4) & 0.321(10) & 1.35 & 0.75(4) & 7(6) & 3(3) & 0.40(29) & 0.15(10) \\
Experiment~E1 & 0.164(3) & 0.143(4) & 0.039(1) & $\cdots$ & 1.18 & 0.70(4) & 12(9) & 8(5) & 0.19(16) & 0.08(7) \\
Experiment~F1 & 0.028(2) & 0.129(1) & $\cdots$  & $\cdots$ & 1.03 & 0.32(3) & 37(26) & $\cdots$ & 0.38(20) & $\cdots$ \\
Experiment~F2 & 0.085(3) & 0.152(8) & 0.021(1) & $\cdots$ & 1.10 & 0.44(3) & 25(20) & 15(8) & 0.32(23) & 0.09(8)\\
Experiment~F3 & 0.138(6) & 0.223(3) & 0.145(5) & 0.070(4) & 1.18  & 0.65(5) & 24(13) & 12(9) & 0.19(15) & 0.06(5)\\
\hline
40\% A3 & 0.221(2) & 0.289(3) & 0.401(4) & 0.495(4) & 1.36  & 0.41(3) & 20(17) & 9(9) & 0.11(8)  & 0.06(5) \\
30\% A4 & 0.218(2) & 0.338(3) & 0.410(4) & 0.495(4) & 1.39 & 0.57(4) & 31(28) & 3(3) & 0.16(13) & 0.05(5)  \\
40\% C1 & 0.216(2) & 0.290(3) & 0.407(4) & 0.495(4) & 1.36  & 0.50(4) & 14(11) & 5(5) & 0.12(9)  & 0.06(5)  \\
30\% D3 & 0.203(2) & 0.334(3) & 0.410(4) & 0.495(4) & 1.36  & 1.13(7) & 7(6) & 3(3) & 0.09(6)  & 0.06(5)  \\
30\% F1 & 0.227(2) & 0.334(3) & 0.402(4) & 0.495(4) & 1.39 & 0.52(4) & 9(8) & 3(3)  & 0.10(7)  & 0.05(4) \\
40\% F2 & 0.224(2) & 0.309(3) & 0.375(4) & 0.493(4) & 1.37 & 0.43(4) & 14(13) & 3(3) & 0.30(25) &  0.06(5) \\
20\% C1/ 80\% C4 & 0.212(3) & 0.214(2) & 0.259(5) & 0.317(12) & 1.28  & 0.40(3) & 12(9) & 7(6) & 0.09(8) & 0.07(6) \\
\hline
20\% A4/20\% D3 & 0.180(15) & 0.335(3) & 0.410(4) & 0.495(4) & 1.33 & 1.00(7)  & 20(19) & 3(3) & 0.15(12) & 0.10(9)\\
22\% F1/11\% F3 & 0.240(2) & 0.328(3) & 0.373(4) & 0.487(5) & 1.39 & 0.40(3) & 19(15) & 3(3) & 0.25(22) & 0.05(4) \\
\hline
\end{tabular}
\end{minipage}
\end{table*}

\subsection{Do we require too many perturbers?}
\label{2Many}

The above results suggest that, with a sufficiently high rate of perturbing
systems, it seems possible to explain the excess of single tranet systems with
respect to unperturbed systems, and possibly even the period distribution as well.
 However, is the frequency of perturbers that is required beyond observed constraints?

An important issue that matters in this comparison
 is whether the false positive incidence is similar in both the single and multiple tranet
samples. It has been discussed at length that the false positive rate is low in the
Kepler multi-tranet systems (Lissauer et al. 2011a; Fabrycky et al. 2014), but the
same reasoning cannot be applied to the single tranet systems.
 In the case of manual candidate selection, there may be a bias towards
finding more planets in multiple systems due to increased scrutiny, but the latest Kepler sample is supposedly selected
entirely automatically (Coughlin et al. 2016). If the false positive rate amongst the single
tranet candidate systems is $\sim 50\%$, then this could explain the discrepancy in $f(50|10)$
between single and multiple tranets without invoking any kind of perturbation.

However, there have been multiple attempts to constrain the false positive rate amongst
Kepler candidates. Radial velocity follow-up for
giant planet candidates (Santerne et al. 2012) find false positive
rates approaching this value, but a variety of statistical studies (Morton \& Johnson 2011;
Fressin et al. 2013; Coughlin et al. 2014), and Spitzer
confirmation studies (Desert et al. 2015) all suggest that the false positive rate for candidates
with $R < 4 R_{\oplus}$ is of the order of 15\% or less, which is too small to account for the
observed single tranet excess. Only Col\'{o}n et al. (2012) find a 50\% false positive rate in their
study of 4 short period (P$< 6$ days) Kepler candidates, but their sample is largely excluded
by our minimum period cut.
 Perhaps the most direct statement can be made using the data from Desert
et al. (2015), who observed 51 Kepler candidates with Spitzer to verify that the transits were
achromatic, as expected from the planetary interpretation. Their choice of targets spanned a range of
parameter space, and 13 match our requirement of being single tranet systems, with orbital periods
between 4--260~days and $R< 4 R_{\oplus}$. Five of these have been directly confirmed and none showed
a substantial false positive probability, whereas a 50\% false positive rate would imply $\sim 6$--7
failures. As such, we discount the possibility that this excess represents a signal from a substantial
population of false positives.

A variety of authors have tried to estimate the occurrence rate of planet
detections around Sun-like stars (Borucki et al. 2011; Catanzarite \& Shao 2011;
Youdin 2011; Howard et al. 2012; Dong \& Zhu 2013, Fressin et al. 2013).
We focus here on the analyses of Christiansen et al. (2015) and Petigura, Howard
\& Marcy (2015) because these authors have done the most comprehensive estimates of
detection efficiency.

In the case of Christiansen et al. -- C15 -- the authors performed
detailed injection and recovery tests in the Kepler pipeline for Kepler
Q9--Q12 data, and characterise the detection efficiency of each Kepler
planet candidate orbiting FGK stars. If we adopt their results for
the period range 5--160~days, we find a total occurrence rate
of $37 \pm 13\%$, for planets whose radii are $1 < R/R_{\oplus} < 2$.
This is calculated on a planet-by-planet basis, so the occurrence rate
of planetary {\em systems} will be lower, depending on the average
number of tranets per star $<N>$. For instance, if all the underlying systems
were generated by our unperturbed model, $<N>=1.55$ and so the true
frequency of planetary systems around FGK stars would be $24 \pm 8$.

Petigura et al. -- PHM -- perform an independent analysis of the Kepler data set,
and perform their own injection and recovery tests to quantify the incompleteness.
Over the same radius range as C15, and covering the
period regime 6.25--200~days, they find an occurrence rate of $28 \pm 6 \%$.
In this case, they only consider the highest SNR tranet for each star, so this
is a more direct measure of the frequency of planetary {\em systems}. Nominally then,
one could infer $<N> = 37/28 = 1.3 \pm 0.5$ from the comparison, although this 
procedure is likely to be affected by the differing systematics. Nevertheless
it agrees with the estimate of $1.1 \pm 0.3$ from Youdin (2011).

Given these empirically estimated occurrence rates, we can predict, for a given
model of the underlying planetary population, what occurrence rate of perturbers
we require. In the simplest case where a single experiment result matches the
observed $f_2/f_1$ value, we might then predict an overall frequency from
the C15 result by dividing by the corresponding $<N>$. For Experiment~A1, this
would imply an overall frequency of $31 \pm 11$\% and, for Experiment~D1, we 
would predict $30 \pm 11$\%. These are both consistent with the PHM estimate,
but underpredict the higher multiplicity tranet systems.

A more plausible estimate comes from the various combinations given
in Table~\ref{tab:TranTab} and Figure~\ref{fig:Rat5}. In the case of the 
best fit model from Figure~\ref{fig:Rat5}, which contains 40\% from Experiment~F2
and 60\% unperturbed systems, $<N>=1.37$ and thus the C15 estimate
corresponds to $27\pm 10\%$, in excellent agreement with the PHM 
estimate. In this case, $f_J=$40\% of all systems with tranets have perturbers
in longer period orbits. We also have to account for the fact that 5/50 of
the Experiment~F2 runs yielded no surviving planets interior to 1~AU and
so will not be included in the above estimate, but will contribute to
a potentially observable populaton of perturbers. As such, our fit corresponds
to a true perturber frequency of $f_J=44\%$. Combining this with the empirical
occurrence rates, 
we expect an overall frequency of $12 \pm 4\%$ from
C15, or $12 \pm 2\%$ from PHM. This corresponds to candidates with 
$R < 2 R_{\oplus}$. 
We get very similar results for the case of 30\% D3 ($<N>=1.38$).
or 30\% A4 ($<N>=1.39$). 
Perhaps the most realistic estimate is to combine the results of
Experiments F1 and F3 to represent a perturber population with
a range of inner perturber locations. If we select a perturber
population that is 2/3 from F1 and 1/3 from F3, and make this 1/3
of all systems, then we expect a population that has 22\% of
systems with perturbers at 2~AU and 11\% with perturbers at
5~AU. However, these experiments are also the most dynamically
excited, and so we need to increase the frequency by a 
factor of 1.47 for F1 and 1.25 for F3 (to account for systems
with no surviving inner planets).
If we restrict ourselves to the
$R < 2 R_{\oplus}$ planets, we thus require $\sim 9 \pm 5\%$ from
F1 and $4 \pm 1\%$ from F3, for a total of $13 \pm 5\%$.

The low mass planet population is not restricted to $R < 2 R_{\oplus}$ however.
If we extend our sample definition to include larger radii planets, the occurrence
rates also increase. PHM extended their analysis up to $R = 4 R_{\oplus}$.
So, if we assume that the planets with radii from 2--4~$R_{\oplus}$ have similar origins to those
with $R < 2 R_{\oplus}$, perhaps with just a small additional Hydrogen envelope, then
the total occurrence rate over the same period range increases to $59 \pm 11 \%$.
Thus, if we repeat our fit of the model with $f_J=$40\% from Experiment~F2, then we
require a frequency of $26 \pm 5 \%$ of all stars
 be orbited by perturbing giant planets. Alternatively, our model constructed
from F1 and F3 implies frequencies of 
 $\sim 19 \pm 4\%$ and $8 \pm 2\%$ respectively.

Other possible mixtures (40\% A3 or 30\% A4) yield very similar
estimates, since $<N>$=1.36 and 1.39 in these two cases as well.
One important difference in these two cases is that the perturbers are
now small enough to potentially avoid constraints from radial velocity
surveys, which makes the estimated companion frequency harder to
constrain.

Despite the potential variations in perturber populations, the requirement
that the combination of perturbed and unperturbed populations match the
tranet multiplicity statistics enforces a robust estimate of the needed
occurrence frequency of such perturbing planets. If we base our projections purely
on the sample of planets with $R < 2 R_{\oplus}$, then we require that
$\sim 11 \pm 4\%$ of FGK stars must possess giant planets on scales
of $\sim$1--5~AU. If we require that planetary systems containing
planets up to $4 R_{\oplus}$ are sculpted by this process, then our estimates
lie closer to $24 \pm 4\%$.

\subsubsection{Radial Velocity Constraints}
\label{RV}

Such perturbers are potentially already constrained by existing and ongoing
surveys. Simple estimates of the yields from radial velocity surveys
given in Udry \& Santos (2007) suggest that the Coralie, Keck and ELODIE
surveys all yield rates $\sim$5--7\% of search targets. More sophisticated
analyses including selection effects have been performed by Cumming et al. (2008)
and Mayor et al. (2011).

Our interest here is particularly in perturbers with semi-major
axes $\sim$ 1--5~AU. Systems are known with giant mass planets
interior to 1~AU along with a handful of companions (55~Canc, WASP-47)
but these are a minority and we assume that such systems do not
contribute substantially to the occurrence rates discussed here. 
As we have shown, planets with larger semi-major axis can still have
an influence, but it is weaker and such perturbers are not likely
to produce enough low tranet systems to match the observations.

From Table~1 of Cumming et al. (2008), we find a frequency of
$8.9 \pm 1.4\%$ for planets with masses 1--15$M_J$ and periods 
less than 11.2~years (corresponding to a semi-major axis of 5~AU), as
measured by the Keck radial velocity survey.
We subtract the frequency of $1.9 \pm 0.6$\% for corresponding
planets interior to 1~AU, leaving us with a frequency of $7.0 \pm 1.4$\%.
If we lower the mass limit to $0.3 M_J$, we find an additional
$1.8 \pm 1.3\%$ of planets between 1--3~AU (detection limits restrict the
sample within 3~AU). Thus, the frequency corresponding to classes
of perturbation considered in the above experiments is $\sim 8.8 \pm 1.4\%$.
This is, of course, a lower limit and indeed, 
  Cumming et al.
fit a power law distribution to the known data and extrapolate that up to
20\% of stars may have gas giants if one extends the range out to 20~AU.

A similar analysis of the HARPS/CORALIE survey by Mayor et al. (2011) yields
a frequency of $13.9 \pm 7$\% for planets with mass $>0.15 M_J$ and period
less than 10~years. This is comparable to the $8.8 + 3.7 = 12.5$\% we infer
from the corresponding numbers in Cumming et al.

Thus, if we consider the acceptable range for inner perturbers to be
5~AU, we may take the frequency of Jupiter mass perturbers to be
$\sim 10 \pm 2\%$. This matches well the required abundance of perturbers
if we restrict our attention to the frequency of short period planets
with $R < 2 R_{\oplus}$. If we wish to explain the frequency of planets up to
$R < 4 R_{\oplus}$, then the RV census appears to fall short by a factor
of two. However, we have shown that a population of sub-Saturn mass planets,
which can evade RV detection, can also provide sufficient perturbations if
it is dynamically hot enough. This could fill in the gap if the demographics
are comparable to that of the more massive planets.

Another important feature of our model is that the largest effects come
from systems with {\em multiple} giant planets, so the effectiveness of
this model also depends on the fraction of giant planet companions with
companions. Bryan et al. (2016) have analysed the frequency with which
known giant planets are accompanied by radial velocity trends suggesting
more distant companions. They find that, for planets with $M > 0.5 M_J$
and $a > 1$AU, the frequency of additional companions on scales $> 5$AU
is at least $54 \pm 7$\%, suggesting that a lot of the known potential
perturbers are in multiple systems.

\subsubsection{Long Period Transits}

In principle, the Kepler data can be used to constrain the frequency of
perturbers directly. The probability of a transit drops with increasing
semi-major axis, but the large number of Kepler observations implies
a non-zero sample of transitting planets at $> 1$~AU scales nevertheless.

For each perturber scenario, we can estimate the probability of observing
the perturbers in transit as well as interior tranets. In the case of 
Experiment~F2, we find that the expected number of single tranets
due to perturbers should be $3.2 \pm 0.3 \%$ that of the number of
single tranets due to smaller, interior planets. Furthermore, a fraction
of the total long period transits should also occur with one or more
tranets at shorter periods. For this scenario, 
 there should
be an increase in the total number of long period tranets by a factor 1.22
relative to single, long-period tranets. Therefore, we may predict that for the case
in which 40\% of systems are perturbed in this fashion (which fits the
multiplicity statistics in Figure~\ref{fig:Rat5} best), $17 \pm 2$ single
long-period tranets ($21 \pm 3$ total) should be found, based on the 1115 single tranets
in our observational data set.

In the case where the perturber is smaller, but closer, like in Experiment~A4, the
probability increases. For the scenario which contains 30\% of A4 in addition to
unperturbed systems, the overall expected frequency of transitting perturbers is 
$2.6\% \pm 0.2\%$ of all single tranets, yielding an expectation of $32 \pm 2$
in the Kepler sample. The greater inclination dispersion results in fewer in
multi-tranet systems, only 4/32. 

As another example, we
 can also estimate the expected number in the case where our perturbing
population contributes 1/3 of all systems and is split between Experiment~F1
and Experiment~F3, as described in \S~\ref{2Many}. In this case, the contribution
from Experiment~F3 is poorly sampled because the orbital periods are long
compared to the Kepler mission lifetime. We find that $\sim 30\%$ of all
single tranets are drawn from Experiment~F1, but that accounting for those
perturber systems which remove all planets raises the equivalent number to
44\%. The expected frequency of perturbers is almost identical to that in
Experiment~F2 and so we find a similar expectation value. Although the 
Experiment~F1 results make a smaller contribution to the observed tranet
population, the higher rate of excitation means more systems are completely
emptied of planets, bringing the required underlying population of perturbers
to the same value once the results are normalised to the observed tranet ratio.

Although the nominal Kepler detection threshold for tranet detection requires
three transits during the mission lifetime, there have been several recent
efforts to systematically search for, and verify, candidate systems with only
single transits in the Kepler database (Wang et al. 2015; Uehara et al. 2016;
Foreman-Mackey et al. 2016). These searches have produced a number of candidates
and suggest that the overall occurrence frequency of these objects is quite
high. The most systematic attempt to measure completeness and exclude false
positives is that of Foreman-Mackey, who find a total of 16 single transit
tranets from a search of 39036 Sun-like Kepler stars. 
This is encouragingly close to
the fraction we expect, although one must be cautious about different systematics
between the Foreman-Mackey search and the nominal Kepler pipeline.
Foreman-Mackey et al also find 5/16 = $31 \pm 14 \%$ of these
planets are in systems with additional transitting planets, which is again in
good agreement with our expectations.
Furthermore, their Table~6
demonstrates that $\sim 80\%$ of this population is composed of planets
with $R < 0.4 R_J$, suggesting that the sub-Jovian population is the dominant
contributor. Such a split may also make this claim consistent with the constraints
based on radial velocity searches.

We conclude that there is indeed evidence in the Kepler data for a population of perturbers at long periods
that may be sufficient to explain the single tranet excess, and that the observational results are
consistent with the notion that
approximately  half the population
is composed of genuine giant planets and the other half may be
composed of sub-Saturn planets. This population would also need to be 
sufficiently dynamically hot to perturb the interior systems sufficiently.
Such a population may also help to address the shortfall found in \S~\ref{RV},
where the RV census of Jupiter-mass planets appeared to fall short by a similar factor.

 As a test of the influence of such a mixed population, we include in
Table~\ref{tab:TranTab} the properties of a model population that is
composed of 60\% unperturbed systems, 20\% drawn from Experiment~A4 and
20\% from Experiment~D3. The combination provides a decent match
to the transit multiplicities and the contribution from Experiment~D3 also
leads to an improved value of $f(50|10)=1.00 \pm 0.07$, although it
still falls a little short of the observations. Such a model also has a better
chance of meeting the requirements of radial velocity surveys, as it
would require only that $12 \pm 2\%$ of FGK stars host Jupiter-scale
companions to match the $59 \pm 11\%$ overall occurrence rate from
PHM.

\subsection{Obliquities}

As noted before, one way to avoid the constraints from radial velocity
observations is to invoke perturbers that are too small to be detected
in that way. However, in order for such perturbers to exert a substantial
effect, they need to have substantial amplitudes. In particular, we have
investigated the influence of $0.1 M_J$ planets with high inclinations.
These, in turn, leave surviving planets on highly inclined orbits.

\begin{figure}
\includegraphics[width=\columnwidth]{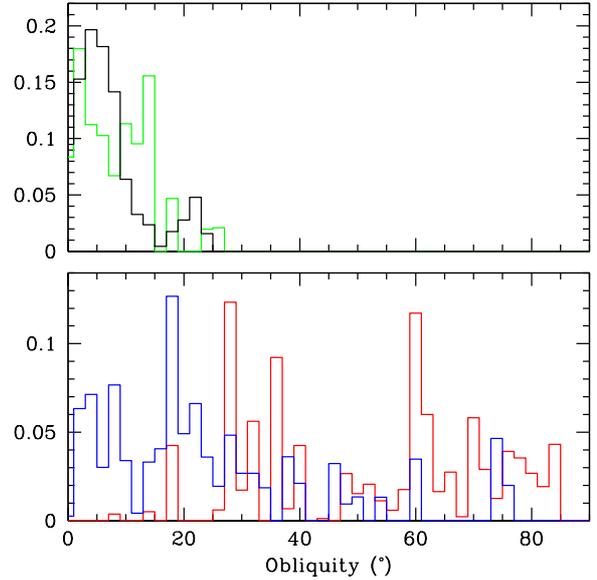}
\caption{The black histogram shows the distribution of obliquities
for the single tranets drawn from our unperturbed planetary systems. The green
histogram shows the result from Experiment~D3, in which the eccentricities are
pumped by approximately coplanar perturbers. The blue
histogram shows the obliquities for those perturbed as in Experiment~F2, where
the perturbers show a range of eccentricities and inclinations. The red histogram
corresponds to the output from Experiment~A4, in which the perturber is inclined
at $60^{\circ}$ but of mass $0.1 M_J$.
\label{fig:Stilts}}
\end{figure}

The difference between eccentricity pumping and high inclination perturbers
 may  be testable by studying the obliquities of Kepler
planet candidates.
 Figure~\ref{fig:Stilts} shows the inclination of the single
tranets observed in  four model populations that illustrate the consequences
of different mechanisms of excitation. In the upper panel, the black histogram is the obliquity
distribution (defined as the inclination of the orbital plane relative to the
original planetesimal disk plane, assumed to represent the stellar spin equator)
of the original, unperturbed, planet population. We see that the great majority
are found within 15$^{\circ}$ of the original orbital plane. The green histogram in
the same panel shows the distribution that results from Experiment~D3, in which the
perturbers drive strong eccentricity growth, but are largely coplanar. We see that
only a small shift in the obliquities is to be expected in this event.

On the other hand, the lower panel shows two populations with greater inclination
excitation.
 The blue histogram
shows the distribution for single tranets that result from Experiment~F2. This
is one of the more promising sources for raising the number of single tranets
observed and achieving agreement with the observations (see Figure~\ref{fig:Rat5}).
In this case we have dynamical heating from both eccentricity and inclination,
and so we do observe some broadening of the obliquity distribution.
The red histogram shows the results of
Experiment~A4, in which orbits are perturbed by a $0.1 M_J$ planet inclined at
60$^{\circ}$ with respect to the original plane. We see that the resulting
dynamical instability leaves a wide range of obliquities, including a prominent
peak at the perturber value and values that
approach polar orbits. Thus, we conclude that a measurement of the
single tranet obliquity distribution can, in principle, indicate that planetary
systems are being perturbed by a population of bodies that are highly inclined,
even if they are too small to be detected in radial velocity searches.

Observations of the obliquities of small planets ($R<4 R_{\oplus}$) in both single and
multiple planets are accumulating (Sanchis-Ojeda et al. 2012; Hirano et al. 2012, 2014; Albrecht et al. 2013;
Chaplin et al. 2013). Evidence to date suggests their orbits are much closer to coplanar than those of 
short period giant planets, but there are hints that single tranet systems may have a larger dispersion
in obliquities than multiple tranet systems (Morton \& Winn 2014). This would argue in favour of the
model presented here and could potentially even offer the possibility of distinguishing between possibilities
as the data improve.

\subsection{Period Distribution}
 We noted in Figure~\ref{fig:Pdis_bin} that the
observations indicate broadly similar period distributions for tranets in single and multiple
systems, while the unperturbed models show a much greater concentration of singles at orbital
periods $<10$~days. Thus, it is of interest to examine the period distribution of the surviving
planets in perturbed systems. We have assessed the various scenarios in terms of the statistic
$f(50|10)$, which measures the ratio of the single tranets observed between 10--50 days and
those between 4--10 days.

\begin{figure}
\includegraphics[width=\columnwidth]{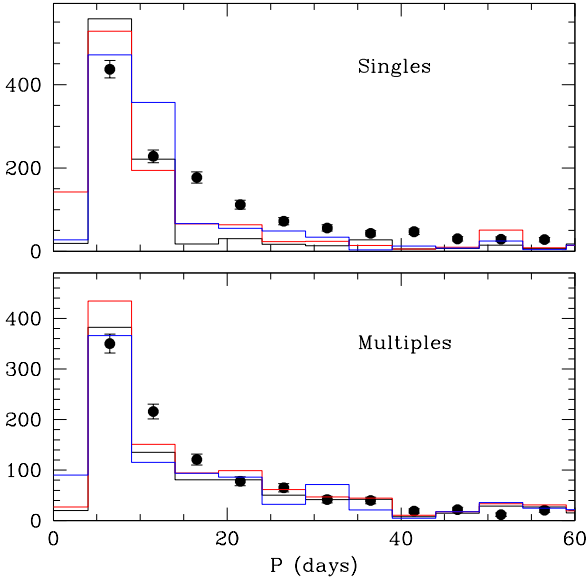}
\caption{The points in the lower (upper) diagram show the period distribution
of observed Kepler tranets in multiple (single) tranet systems. The black histogram shows the
model population constructed with 50\% of systems drawn from Experiment~F2, in which the planetary systems were perturbed
by a pair of Jovian mass planet with eccentricities and inclinations drawn from a diluted version of the distribution of 
Juric \& Tremaine (2008).
 The red histogram
shows the effects of a population in which 40\% of systems are drawn from experiment A4,
 in which the perturber is a $0.1 M_J$ planet on a highly inclined orbit (60$^{\circ}$).
 Finally the blue histogram shows a model population composed of 70\%
unperturbed systems and 30\% drawn from Experiment~D3, in which the perturbations arise from
a giant planet pair with significant eccentricity, which drives secular instability and scattering,
and which comes closest to matching the observed period distribution.
\label{fig:Pdis_bin4}}
\end{figure}

Figure~\ref{fig:Pdis_bin4} shows the period distributions of single and multiple tranets for
 three models discussed in \S~\ref{TwoPop}, chosen to roughly match the
tranet period ratios. The black histogram represents the population that is drawn
50\% from Experiment~F2 and 50\% from the unperturbed population (so, the same model
as the solid histogram in Figure~\ref{fig:Rat5}). We see that it still suffers from
the same problem as most of our models, in that it produces a value of 
$f(50|10)$ that is too low, i.e. too many single tranets at short periods.
The blue histogram represents our best case in this regard, namely a population
that is 30\% drawn from Experiment~D3 and  70\% from the unperturbed population.
We see that the pile-up of single tranets in the range 10--15~days improves the
$f(50|10)$ statistic, but still underpredicts the counts at longer periods.
 The red histogram represents the population that is drawn from
Experiment~A4 in 40\% of cases, and from the unperturbed population in the other 60\%.
This population does slightly better than the black histogram but not as good as the
blue. However, the perturbers in this case would be difficult to find with radial
velocities. 

In summary, it appears as though the period distribution of single
tranets likely contains information regarding the distribution of separations of
perturbing pairs, favouring closer pairs.

\begin{figure}
\includegraphics[width=\columnwidth]{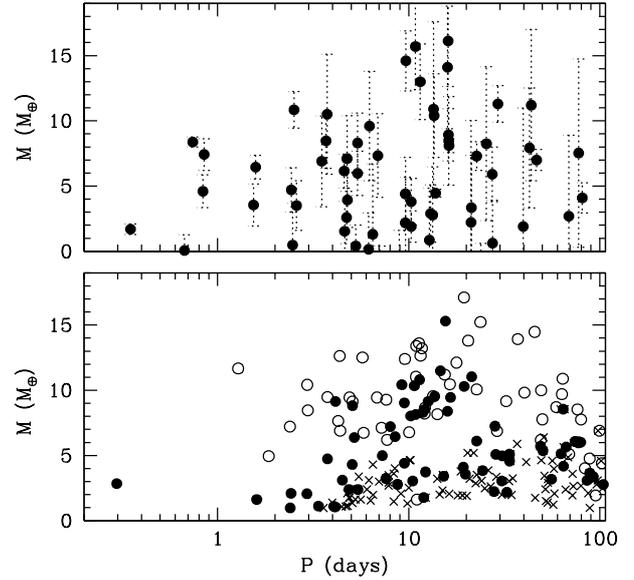}
\caption{The upper panel shows the full observed sample of masses and periods from
the compilation by Weiss \& Marcy (2014). The bottom panel shows the corresponding distribution 
for three model populations. The crosses indicate the planets in systems with no external perturbers.
The filled circles are the outcome of Experiment~D1, in which the systems are pumped by a pair of
distant Jupiter-mass planets on coplanar but eccentric orbits. The open circles correspond to the 
results of Experiment~A4, in which the perturber is only $0.1 M_J$ but inclined by $60^{\circ}$ to
the orbital plane of the original planetesimal disk. In both of these latter cases, we have accounted
for the evolution due to the combination of tidal dissipation and secular coupling, as described
in HM15 and in the text.
However, we have not accounted for the possibility that some of the planets excited to sun-grazing
orbits could have been captured into short-period orbits.
\label{fig:MRcomp}}
\end{figure}

Another way to probe the possible signatures of secular perturbations is to examine the
parameter space of surviving systems to identify unique features. Figure~\ref{fig:MRcomp} shows
the period-mass distribution of observed systems from Weiss  \& Marcy (2014), compared to
that from a handful of simulations. The crosses in the lower panel show the output from the
original, unperturbed simulations. Filled circles show the results of Experiment~D1 after we
have also incorporated the effects of tidal evolution in our secular model (\S~\ref{TidEv}) and
the open circles indicate the same for Experiment~F2. 

The observed sample is drawn in a heterogenous manner, including systems with Neptune or
Jupiter size planets on much smaller scales than we have assumed here (e.g. 55~Canc~e -- Fischer et al. 2008
or Kepler-18b -- Cochran et al. 2011), and so population comparisons with models are not reasonable.
Nevertheless, we can compare classes of planet to see which are realised in different Experiments. In those parts of parameter space where
circles and crosses overlap, we cannot easily distinguish the signatures of secular perturbations, because both unperturbed
and perturbed models can produce such planets.
However, for periods $< 3$~days, or for $M > 5 M_{\oplus}$ and $P < 20$~days, we find that our
unperturbed model produces very few planets, while the perturbed systems populate these regions.
In particular, the group of planets anticipated with $M \sim 10 M_{\oplus}$ and $P \sim 10$~days
appear to have an observed counterpart. It is these latter systems that we expect to have a significant
fraction of distant perturbers, although it is possible that multiple processes contributed (We have
ourselves argued that 55~Canc~e is the result of Roche-lobe overflow of a Neptune-mass object-- Hansen \& Zink 2015).

\subsection{Period Ratios}
\label{Prats}

In \S~\ref{MoDis}, we found that proximity to second order resonance played a protective role
in cases where the perturbers were sufficiently distant. This prompts the idea that perhaps
the excess of near-resonant pairs identified in the Kepler data (e.g. Lissauer et al. 2011a;
Fabrycky et al. 2014) is a consequence of such configurations being more resilient to
secular perturbations by virtue of their higher precession frequencies (see Malhotra et al. 1989).

\begin{figure}
\includegraphics[width=\columnwidth]{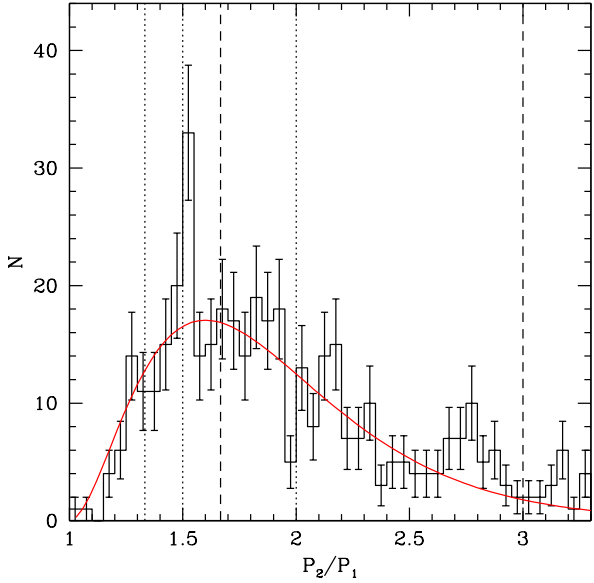}
\caption{The histogram shows the period ratio distribution for the 
Kepler sample outlined by Coughlin et al. (2015), after accounting for $4 < P < 260$~days,
and $T_{eff} > 4000$~K. Vertical dotted lines show the locations of first order resonances
and dashed lines second order resonances. The red curve is a fit used to interpolate
across the resonant features to estimate excesses.
\label{fig:Pbincomp}}
\end{figure}

To assess this, we must
first quantify the excess. Figure~\ref{fig:Pbincomp} shows the distribution of period ratios for
the observational sample as defined in \S~\ref{KSTE}. The two most obvious
features are the spike near the 3:2 resonance and the apparent deficit just inside the 2:1
resonance, as previously noted by Lissauer et al. (2011a) and Fabrycky et al. (2014). To quantify
the excess, we fit a function of the form $ x^2 exp(-x/x_0)$ to the overall distribution
(where $x=P_2/P_1-1$) and $x_0=0.3$. This is shown as the red curve in Figure~\ref{fig:Pbincomp}.
Using this function to interpolate over the range 1--3, we find that 11\% of planet pairs
lie within $1.45<P_2/P_1<1.55$, whereas the smoothed distribution predicts 7\%. In the
case of the range $1.95<P_2/P_1<2.05$, the observations show 4\% whereas the smoothed distribution
predicts 5\%. Thus, in truth, the excess near the 3:2 resonance represents about 4\% of all
planet pairs, and there is very little difference in overall frequency near the 2:1 resonance --
although the distribution indeed appears asymmetric (the evident
dip is partially cancelled by a slight excess at larger values). 

The unperturbed simulations show 4\% of systems near (using the same period bin as for the observations) the 3:2 resonance  after accounting for
the transit observation selection effects and 8\% near the 2:1 resonance. We can now repeat the exercise with the outputs of
the various experiments. In no case do we find a significant increase. The fraction of pairs near the 3:2 resonance
ranges from 3.1\% (Experiment~C4) to almost zero (e.g. Experiment~D1). Similarly the fraction near the 2:1 resonance
remains between 7 and 8\% for weaker perturbations (Experiments C2 \& C3) but again drops to about 3\% in the
case of stronger perturbations (Experiments~A4, D1, F1, F2, F3). In essence we find no evidence in our model for a significant
increase of survival for near-resonant neighbouring pairs.
This failure to reproduce the observational signature
stems from the fact that dynamical instabilities play a substantial role in the Experiments that fit the data, and planet-planet
scattering and collision overrides subtleties in the secular architecture.

This may seem at odds with the results of \S~\ref{MoDis}, which found that resonant effects (albeit mostly
second-order) seemed to play a role in protecting some systems from secular instability. However, we note that these
effects occur on scales $\sim 1$AU and so are severely under-represented in a sample observed in transits. We 
also found that the perturbations in that Experiment were too weak to explain the observations. 

\subsection{Signatures of Planet Collisions?}

Another potentially observable consequence of secular perturbations is its effect on the
planets themselves. Orbital instability is associated with scattering and planetary collisions, which
may result in differences between the observable properties of planets that suffered late-time collisions
and those that did not. If we consider the results of experiment~F2, we find that the evolution of 50
systems yields a total of 77 planet collisions, 59 collisions of a planet with the star and 9 ejections.
Of the 86 surviving non-Jovian planets, 33 suffered at least one collision (38 \%). The incidence of collisions
is higher for short period survivors, and so, when we weight the fraction by transit probability, we
find that we expect 64\% of single tranets derived from Experiment~F2 to have experienced at least one
collision with another planet. Therefore, in population defined by a mixture of Experiment~F2 and unperturbed
systems, we might anticipate that the single tranet population would be composed of a higher fraction
($\sim 40\%$) of
tranets that have experienced a late collision.
 One of the surprises of the low mass planet population
is the number of planets with low densities (Lissauer et al. 2011b; Wu \& Lithwick 2013), suggesting low density components comprising 1-10\% of the
mass. If this material is that which is most easily lost during collisions, we might expect the single tranet
population to have fewer low density members (see Inamdar \& Schlichting 2015 for a related discussion concerning
the effects of late giant impacts during the assembly process).

\begin{figure}
\includegraphics[width=\columnwidth]{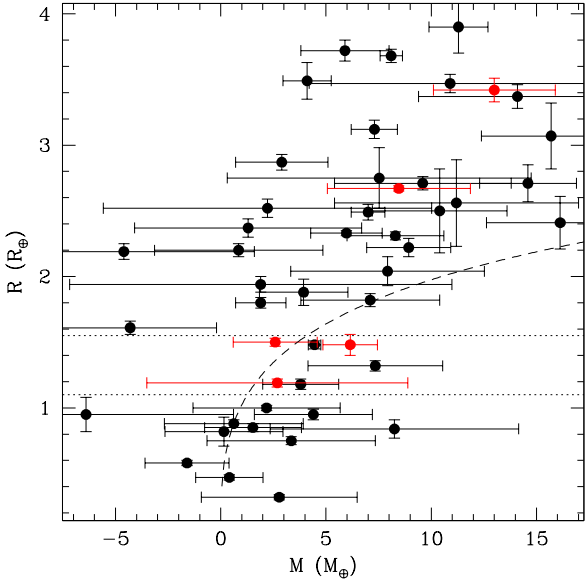}
\caption{The measurements of mass radius shown here are taken from Weiss \& Marcy (2014).
Black points show planets in multiple tranet systems and red points show those in single tranet
systems. Only planets with orbital periods $>4$~days are plotted. The dashed curve is the mass
radius relation corresponding to the Perovskite equation of state from Seager et al. (2007). The
horizontal dotted lines denote a radius range from 1.1--1.55$R_{\oplus}$.
\label{fig:MR}}
\end{figure}

Figure~\ref{fig:MR} shows the mass-radius measurements tabulated by Weiss \& Marcy (2014) as the result
of an ongoing program to measure radial velocities for Kepler planet candidates. We have
limited the comparison to orbital periods $>$4~days to match the model described here, and
plot different colours depending on whether the planet is in a single (red) or multiple (black)
tranet system. There are only five planets in this set that correspond to single tranets and
so it is difficult to draw any substantial conclusions. However, it is notable that 3/5 of the
single tranets have $1.1 R_{\oplus} < R < 1.55 R_{\oplus}$ while only 3/42 of those in multiple
systems do. We choose this radius range under the assumption that the products of planetary collisions
will be more massive than average but will also have lost any low density component. Thus, we
anticipate they will favour the upper end of the mass-radius relation for rocky planets, but should
not exceed it. We urge more radial velocity measurements of single tranet systems to examine
whether this tantalising but statistically marginal difference between single and multiple
tranet systems is real.

\subsection{Signatures of Stellar Collisions?}

One of the striking aspects of the simulations shown here is the high rate of stellar collisions,
which contributes to the reduction of multiplicity at a rate that is non-negligible relative
to planetary collisions, and well above that due to planetary ejections.
 As we discuss in
\S~\ref{MultiPerte} and \S~\ref{RLO} , some of these systems may not be
absorbed or disrupted and may be tidally captured into
extremely short period orbits, possibly matching a population emerging from observations.
Some of them will almost certainly also collide with the star. For the dense, rocky bodies
studied here, Metzger, Giannios \& Spiegel (2012) estimate that the planet will plunge
beneath the stellar photosphere, depositing the orbital energy in a wake, and producing an
EUV or soft X-ray flash. Their estimates were performed for Jupiter-mass planets and so
are likely optimistic for this case in terms of luminosity, but pessimistic in terms of
frequency. A $5 M_{\oplus}$ planet will spiral in due to gas drag in a few orbital times,
so that the energy input will be on a timescale of order a few hours, and comparable to the
orbital binding energy of the planet $\sim 10^{43}$ ergs. A substantial fraction of
planetary systems may experience such an event in their lifetimes.

\begin{figure}
\includegraphics[width=\columnwidth]{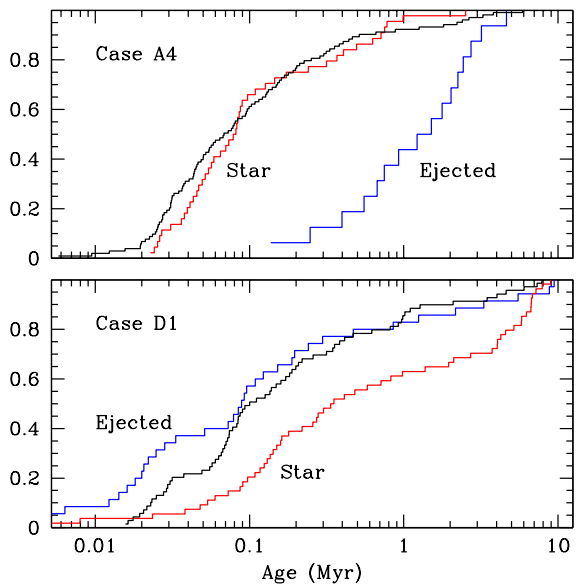}
\caption{The black curve shows the cumulative distribution of system ages at
which a planet-planet collision occurred. The red histogram shows the distribution of ages
at which a planet collided with the host star, and the blue histogram shows the ages at
which planets were ejected from the system. In the upper panel we show the results for
Experiment~A4, in which the perturber was an inclined $0.1 M_J$ planet. In the lower panel
we show Experiment~C1, in which there were two $1 M_J$ planets on eccentric but coplanar
orbits. We show cumulative distributions, which do not reflect the overall frequencies
of the different channels. In the case of Experiment~A4, the frequency of planet~collisions to
stellar~collisions to ejections is 8.6:3.7:1, while, in the case of Experiment~C1, the
same outcomes are found in the proportions 2.0:1.5:1, i.e. much more evenly distributed.
\label{fig:Ages}}
\end{figure}

To estimate an appropriate rate of such events, we need to understand the
characteristic timescale for orbital instabilities to set in.
 Figure~\ref{fig:Ages} shows the
cumulative distributions for the ages at which systems experienced planet-planet collisions,
planet-star collisions, or ejection from the system. We show distributions for two cases that
experienced substantial instability. In the upper panel, we show the results for Experiment~A4, in
which the system is perturbed by a $0.1 M_J$ perturber on a circular orbit inclined by
$60^{\circ}$. We see that the collisions, either with other planets or the host star, occur
with a median system age $\sim 0.1$~Myr and are largely completed by 1~Myr. The time to
eject planets is longer, with a median lifetime $\sim 1.5$~Myr. In the lower panel, we show
the Experiment~C1, in which the perturbation comes from a pair of Jupiter mass planets on coplanar
but eccentric orbits. The distribution of timescales is broader in this case, and the
characteristic timescale for ejections is shorter and more comparable to the planetary
collision timescale. The median time for stellar collisions is now $\sim 0.4$~Myr. These
differences reflect the different manner in which the instabilities are driven. The difference
in ejection distributions is primarily due to the difference in perturber mass -- planetary
scattering by more massive planets is more efficient is more efficient.

The conclusion we draw is that unstable systems experience most of their instability
quite rapidly, and the surviving systems are quite widely spaced and dynamically stable
by the age of 10~Myr. The effects of planetary collisions may reasonably be expected to
produce substantial quantities of dust, but these results suggest that there may be
little measurable delay between planet assembly and instability, except for that potentially
imposed by any offsets between the formation of the interior planets and the perturbing
giant planets. As such, it is not clear that we should expect any discernable interval
between different epochs (assembly and subsequent instability) that show 
 infrared excesses that result from the production of dust. These results also suggest
that the rate of stellar collisions should be $\sim 0.4 \rm Myr/40\% = 10^{-6}$ per year
per star.

As we have noted before, it is possible that tidal interactions may capture planets
from grazing orbits into
short period orbits before they collide with the star. This could provide a mechanism
to population the USP population of planets identified in the Kepler data. Sanchis-Ojeda
et al. (2014) estimate a frequency $\sim 0.5 \pm 0.1\%$ of all G stars possess such a
planet. For our Experiment~F2, we find that 58\% of systems experienced at least one 
planet-star collision. For the model where Experiment~F2 provides 40\% of the underlying
planetary population and that is used to explain the existence of detectable planets
around 57\% of all FGK stars, this implies $\sim 13\%$ of all stars experience at least
one planet collision. Thus, if only 4\% of these events result in surviving planets, that
would match the observed frequency.

In the case where the planets are driven to short period orbits by the combination
of secular excitation and tidal dissipation, the observed frequency of migration
interior to 10 hours period is $\sim 2\%$ in Experiment~D1. The application of
Experiment~D1 to the transit multiplicities requires an unreasonably high frequency
of perturbers, but does produce $f(50|10)$ that matches observations. Thus, it is
also possible that this pathway could contribute to the observed population of USP.
It would also be consistent with the observation of Sanchis-Ojeda et al. (2014) that
a high fraction of USP candidates were in systems that had another candidate with
period less than 50~days.
 
\section{Discussion}
\label{Disc}

Our goal in these calculations has been to explore the effects of secular pumping
of low mass, compact planetary systems by giant planets on larger scales. We consider
the effects of independently driving inclination and eccentricity, as well as the
influence of single perturbers versus pairs of perturbers.

Our first conclusion is that the pumping of inclination alone cannot produce enough
single tranet systems to match observations without driving a significant fraction
of the systems to a state of dynamical instability. Thus, we expect that the
single tranet systems must contain a significant number of intrinsically low
multiplicity systems. This is consistent with attempts to fit the demographics
 to simple distributions(Fang \& Margot 2012; Dong \& Tremaine 2012) as well as with estimates based
on the frequency of transit timing variations (Xie, Wu \& Lithwick 2014).

As a consequence of this conclusion, we also investigated the effect of driving larger inclinations as well
as eccentricity directly, in order to examine how the reduction in planet multiplicity
 manifests itself in the transit statistics.
We then examined the implications
of potentially successful scenarios for other observables.

\subsection{Single Populations}
 Experiments A1, D1 and E1 all yield sufficient reduction in $f_2/f_1$ to match
the observations. These each represent somewhat different perturber populations.
Experiment~A1 represents the pumping of inclination directly by a single inclined
Jupiter mass perturber, Experiment~D1 demonstrates how a pair of Jupiter mass planets
can pump eccentricity, and Experiment~E1 shows the effects of a single Jupiter mass
planet when the eccentricity and inclination are chosen from a distribution chosen
to match the observed properties of Jovian planets on large scales.

These scenarios alone are also all unsatisfactory for multiple reasons. While they match
the ratio of single to double tranet systems, they all underpredict the frequency
of systems of higher tranet multiplicity. This is not surprising, as the level of
dynamical excitation required to satisfy the former constraint naturally works against
the latter. Furthermore, if any of these scenarios were the explanation for the observed
properties, they would imply that nearly every system with planets on short period
orbits possesses a perturber on larger scales, because the value of $f_2/f_1$ from the
model is only slightly below the observed value. We have reviewed estimates of the observed
frequency of potential perturber systems around stars and conclude that the required frequency is in
well in excess of the current constraints from radial velocities.

\subsection{Multiple Populations}
 A more promising route is to consider even stronger perturbations, that reduce
the underlying multiplicity substantially, but only in a fraction of the systems. We
thus posit that the observed planetary population corresponds to a combination of
systems without perturbers, exhibiting small inclination dispersions and high multiplicity,
and  systems of low multiplicity and high excitation, born of the original high multiplicity
systems but denuded by the operation of a population of Jovian mass perturbers.

Table~\ref{tab:TranTab} presents several combinations of this type, utilising particularly
the results of Experiments A3, A4, C1, D3, F1, F2 and F3. The latter five here all offer some variation
on the theme of two external perturbers with a range of eccentricities and inclinations.
These pump sufficient eccentricity into the interior system to drive substantial dynamical
instability. Indeed, some of these systems remove all the interior planets from the system,
with the giant planets as the only survivors.
Several combinations of this type can satisfy the observed tranet multiplicity distribution
(see Figure~\ref{fig:Rat5}). In essence, the high multiplicity systems derive almost entirely
from the unperturbed population, while the single tranet systems consist of a roughly
equal split of truly low multiplicity systems drawn from the perturbed population, and
high multiplicity systems observed at larger inclination.

The models which feature Experiments A3 and A4 are different in that the perturber is a 
single planet of mass $0.1 M_J$ on an inclined orbit. These are designed to achieve
the necessary level of dynamical instability while invoking a perturber of low enough mass
to avoid detection by radial velocity methods. The required inclinations are $> 30^{\circ}$
and so produce substantial obliquities in surviving planets.

 The advantage of these composite models is that they require a smaller fraction of
stars to host giant planets on large scales. It has become clear of late that the frequency
of low mass planetary systems is higher than that of giant planet systems, and so any
model purporting to explain the tranet multiplicities in this way must satisfy the relative frequencies.

The attempt to compare the frequencies of different components requires some assumptions
about how to map the models to the observations. Our models for the compact systems assume
rocky planets and so are most directly comparable to observed planets with $R < 2 R_{\oplus}$.
If we restrict the application of our models to these systems, our discussion in \S~\ref{2Many}
suggests that the frequency of Jupiter-mass perturbers required to match the observations is compatible
with the estimates based on radial velocity surveys. In this event, we leave an approximately
equal-size population of larger planets ($2 R_{\oplus} < R < 4 R_{\oplus}$) unexplained. This 
may represent a population with a different origin, perhaps
corresponding to the inwards migration posited by some authors (Ida \& Lin 2010; Rein 2012; Mordasini et al. 2012; Cossou et al. 2014 ) 
from larger distances. The larger radii would be consistent 
with an origin with material containing  a larger volatile inventory.

 However, there is little evidence to suggest a substantial difference in the period
distribution of planets with $R < 2 R_{\oplus}$
and those with $2 R_{\oplus} < R < 4 R_{\oplus}$ (e.g. Dong \& Zhu 2013), and systems exist with planets from
both classes (e.g. Carter et al. 2012; Gautier et al. 2012).
We might equally well interpret these as
resulting from the same underlying process as the smaller radius cohort, but possibly retaining a small fraction of the
mass in a Hydrogen envelope. The radii can be explained by a sufficiently small mass in
Hydrogen that it need not affect the assembly substantially.

In this event, the frequency of systems is about twice as large as before, and requires
approximately $25 \%$ of stars to possess a planetary component capable of substantially
perturbing the interior planetary systems. This is larger than can reasonably be supported
by the radial velocity observations, although it may be consistent if we count all the giant
planets, even out to 20~AU. Such a scenario could be realised with a relatively dynamic picture of
giant planet evolution -- if most giant planets spent some early portion of their lifetime
on scales $\sim 1$--5~AU, and only later migrated out to more distant locations. Such a 
picture has indeed been proposed for our own Solar System giant planets (Walsh et al. 2011).
In this picture 
the planets Jupiter and Saturn migrated inwards to within 2~AU (the precise nature of the migration
is still a matter of debate -- Brasser et al. 2016), and
the configuration of our own terrestrial planets may have been strongly affected by either secular resonance
sweeping (Ward 1981; Agnor \& Lin 2012; Brasser, Walsh \& Nesvorny 2013) or other effects (Batygin \& Laughlin 2015).
The interaction of Saturn and Jupiter eventually caused them to migrate outwards and eventually the 
current configuration was put in place by planetesimal scattering (Tsiganis et al. 2005). The fact that
our results show the biggest perturbations result from multiple planet systems may suggest a link to
scenarios of this type, although the level of dynamical excitation we invoke may be at odds with the
dissipation to be expected from systems migrating in a gaseous disk.

 An alternative pathway to finding a sufficiently numerous population of perturbers is to make
up the difference between the observed Jovian population and the requirements with planets small enough
to avoid detection in radial velocity surveys. We have shown that this is indeed possible, although it
requires a dynamically hot population. Encouragingly, there are hints of a substantial
population of sub-Jovian planets on larger scales in Kepler data (Dong \& Zhu 2013; Wang et al. 2015; Uehara et al. 2016; Foreman-Mackey et al. 2016).
Such a population of perturbers does leave an imprint in the form of high obliquities for those planets that
survive in these systems. Thus, a scenario like this one predicts that single tranets should have a larger
fraction of systems with large ($ >  20^{\circ}$) obliquities, relative to the high multiplicity systems
(whose obliquities should remain in the single digits). The current state of observation is still somewhat
uncertain, with contradictory claims regarding the incidence of high obliquity (Hirano et al. 2014; Morton \& Winn 2014).

These results suggest that some or perhaps all of the KSTE can be
the result of dynamical excitation due to secular perturbations from giant planets on
scales $> 1$AU. The required frequency of perturbers is potentially consistent with the
constraints from both radial velocity studies and searches for single tranet candidates in
the Kepler database. It is also possible to match the period distribution with particular
choices of perturber systems.

\subsection{Alternative Models}

However, we must note that these calculations do 
 not present a complete evolutionary
history for these systems, and are best described as a sensitivity study for the robustness
of our model for the unperturbed systems under the influence of these external perturbations. 
We have assumed that our models from HM13 represent
a good representation for the underlying, unperturbed planetary systems. These are derived 
from the assumption of in situ assembly, but may also be an acceptable representation of systems
derived from other evolutionary pathways because it is only the final secular architecture that
matters. Potentially more problematic is if the assembly (or migration) takes place in the 
presence of a pre-existing giant planet system, in which case the giant planets may impose
additional regularities not seen here. Thus, the model described here should be considered as
one possible representation of the interactions between compact and distant planetary systems
and should not be considered an exhaustive exploration of the possibilities. Indeed, other
descriptions of these interactions are being actively pursued (Hands \& Alexander 2016;
Mustill, Davies \& Johansen 2016; Reid \& Wyatt 2016). 

Furthermore, ours is not the only proposal to explain the KSTE or Kepler Dichotomy as it
is more widely described. Several authors have proposed that variations in the conditions
for in situ assembly, either in terms of the surface density profile (Moriarty \& Ballard 2015)
or the level of dissipation in the remnant gas disk (Dawson, Lee \& Chiang 2016), could also
result in sufficient variation in the tranet properties to match the observations. Other
mechanisms for tilting a planetary systems, such as enhanced precession due to the oblateness
of young, rapidly rotating stars (Spalding \& Batygin 2016) have also been proposed. Fortunately,
as discussed above, 
there is potentially sufficient ancillary information available (obliquities, period distributions,
companion frequencies) to eventually tease these scenarios
apart. 

\subsection{Context for the Solar System}

\begin{figure}
\includegraphics[width=\columnwidth]{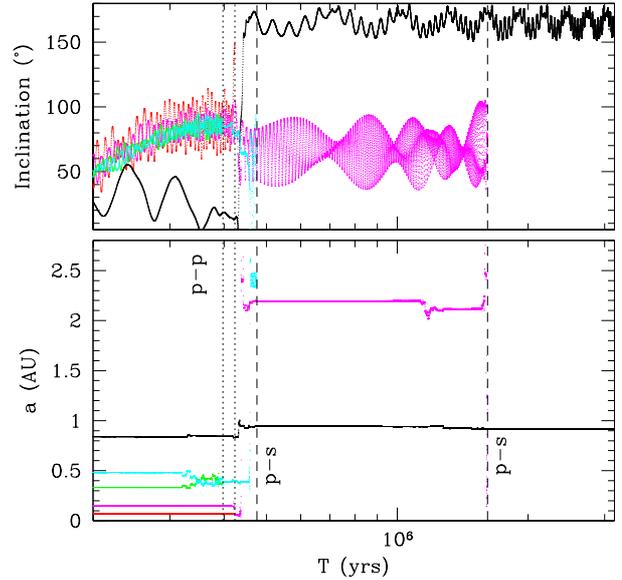}
\caption{The lower panel shows the evolution of the semi-major axis and
the upper panel shows the evolution of the inclination for each planet in the system
(excluding the giant planet perturbers at 5~AU and 24.1~AU).
The black line shows the evolution of  the
planet that survives, while the red, magenta, green and cyan curves show the evolution
of planets that are lost. The two dotted lines indicate the epoch of a planet-planet (p-p)
collision, while the two dashed lines indicate when a particular planet collides with the
star (p-s). In each case the semi-major axis of the impacting planet was outside 1~AU at
the time of collision. The final system is an example of one that shows a large inner
hole, and a surviving Earth equivalent planet ($0.89 M_{\oplus}$ and 0.91~AU) 
but it is not a perfect analogue for the solar system, as the survivor has an eccentricity
of $\sim 0.2$ and orbits in a retrograde sense.
\label{fig:SSex}}
\end{figure}

The substantial diminution of planetary multiplicity seen in many of our simulations
 may also have relevance to our own
Solar System. With the discovery of so many short period planets by Kepler, the solar system
appears to be quite deficient in terms of planets on small scales, since many star now
appear to have multiple planets with periods shorter than that of Mercury. Several authors
have suggested that perhaps we began with more planets but lost them over time. Batygin \&
Laughlin (2015) suggest that our original Solar System may have contained a
population of inner planets that was  swept into the star
by a population of planetesimals scattered inwards by giant planets, captured by the gas
disk and then pulled inwards by drag forces. Alternatively, Volk \& Gladman (2015) suggest
that planetary instability and collision could grind down the planets to dust, leaving Mercury as
the only remnant. The results here suggest that secular perturbations can provide a path that
combines the best elements of both these scenarios. Figure~\ref{fig:SSex} shows an example
of one of our simulations from Experiment~F3, in which an initial five planet system is
whittled down to a single, $0.89 M_{\oplus}$ planet with a semi-major axis of 0.91~AU. 
This system is destabilised by inclination pumping, with the four innermost planets being
tilted coherently to $\sim 90^{\circ}$ within 0.4~Myr, at which point dynamical instability
sets in. The system undergoes two planet-planet collisions and a planet-star collision in
short order, leaving two planets, with the original second innermost on a high eccentricity
($\sim 0.9$) orbit, which is eventually driven into the star on longer timescales.
This leaves the original outermost of the five planet system alone in a potentially
habitable orbit. The final configuration bears some relation to the Solar System in terms of mass and location,
although the final eccentricity is too large ($\sim 0.2$) and the orbit is actually
retrograde ($\sim 153^{\circ}$) -- a consequence of the inclination pumping followed by
scattering. Nevertheless, 
this demonstrates giant planet perturbations can preserve habitable planets while still removing the interior 
system, as in the Batygin \& Laughlin proposal, but without the need to invoke additional
physics associated with coupling of planetesimals to the disk. The role of
stellar collisions is also noteworthy in this regard, as it makes it easier to remove
planets entirely, rather than to simply accumulate them into larger systems. In self-excited
systems such as those posited by Volk \& Gladman, planetary collisions are dominant
and will leave observable remnants unless the collisions are strongly erosive.

\begin{figure}
\includegraphics[width=\columnwidth]{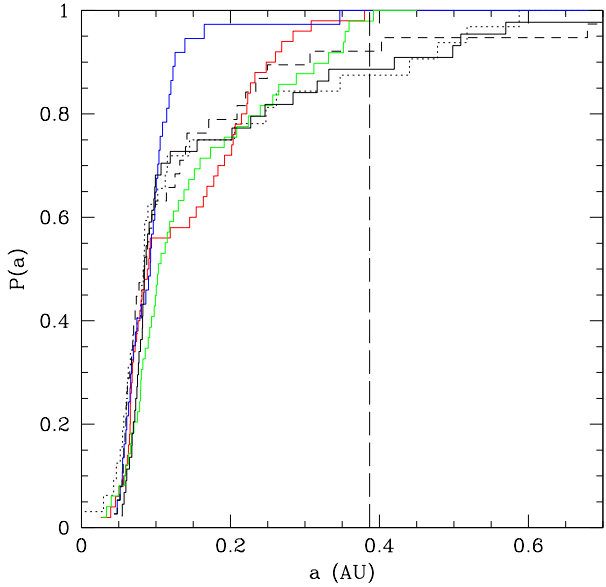}
\caption{Each histogram shows the cumulative distribution of inner planet
semi-major axis for a variety of different scenarios. The red and blue histograms correspond
to Experiments C1 and D1 respectively, while the green histogram represents Experiment~A4. 
The black dotted (dashed) histogram shows the results from Experiment~F1 and F3 respectively, and the black solid
histogram shows the results from Experiment~F2. The vertical long dashed line shows the
semi-major axis of Mercury. We see that only the black histograms shows a substantial
probability ($\sim$5--$15\%$) of producing a system with a hole as big as that in the Solar
System.
\label{fig:GapSS}}
\end{figure}

Figure~\ref{fig:GapSS} shows the probability distributions of the inner planet semi-major axis
for a variety of our perturbation experiments. We see that most of the Experiments, even
those which substantially reduce the tranet multiplicity (like A4 and D1), do not produce
many systems which have surviving planetary systems whose innermost members lie at $a > 0.4$AU.
However, the distributions from  Experiments~F1, F2 and F3
do possess a non-negligible fraction that could potentially provide a pathway to forming the
Solar System terrestrial planets. The fact that these latter distributions are similar suggests
the reason is less a function of the level of dynamical excitation and more the result of having
a range of binary separations and hence forcing frequencies.

Finally, we note that our
 focus here has been on planetary systems orbiting solar-type FGK stars, which dominate
the Kepler sample. However, Ballard \& Johnson (2016)
have noted that a similar single tranet excess can be found around lower mass stars, suggesting that the phenomenon
is ubiquitous. They suggest that the influence of binary stellar companions may be responsible, but,
as we have noted at above, distant companions tend to tilt planetary systems coherently,
rather than increasing their intrinsic dispersion. As such, the similarly of the single tranet
excess around M stars suggests that they too possess distant planetary systems in some abundance.
We aim to repeat the set of experiments performed here using the results from Hansen (2015) in a 
future publication and to compare to estimates of the corresponding populations of perturbers in that case.

\section{Conclusions}
\label{Conc}

We have examined the role of secular forcing on the properties of compact, short-period planetary
systems such as those discovered by the Kepler satellite and radial velocity monitoring campaigns.
The perturbations are the result of giant planets on scales 1--5~AU, along with possible additional
companions on larger scales. We have investigated the effects of perturbing inclination and eccentricity
individually and in combinations consistent with the observed properties of such giant planets.

We find that, if we wish to explain the excess of single tranets observed by Kepler, the level of
forcing required inevitably leads to a lower multiplicity population of planets, even if we start
with a high multiplicity population and perturb only the inclination. This is a consequence of the
amplitude of forcing required, which violates the classical separation of inclination and eccentricity
behaviour.

We find that the measured occurrence rates for small planets detected by Kepler imply a frequency
of perturbers that is consistent with constraints from radial velocity surveys if we restrict our
comparison to planets with $R < 2 R_{\oplus}$. If we apply our model to all planets with $R < 4 R_{\oplus}$
then the required frequency of perturbers exceeds observational constraints by a factor of two. This could still be consistent
with the model if a substantial fraction of the perturbers are of sub-Saturn mass, and thus missing
from radial velocity surveys. Examination of single transit events from Kepler lends support to the
existence of such a population. However, it 
 would need to be dynamically hot to provide sufficient amplitude of perturbation to explain
the observations, and leaves a robust signature of high obliquities in surviving planets. 

A planetary population with such a dynamic history can potentially lead to other observable distinctions.
We have posited that the frequency of late planet collisions amongst the low multiplicity population
should imply fewer low bulk density planets amongst the single tranet population, that the frequent
occurrence of close star-planet interactions could produce a population of very short period planets
and that planetary systems like the inner solar system (with large inner holes) can occur in modest
proportions ($\sim 10\%$ of all perturbed systems and so in $\sim 5\%$ of all cases). Continued radial
velocity follow-up of transitting candidates, extended to multi-year baselines, and including a representative
sample of single transit systems, should help to clarify the links between tranet multiplicity and
distant giant planets. A continued effort to determine planetary obliquities will also clarify the
excitation level of the surviving systems and shed light on potential perturbers.

\section*{Acknowledgements}

This research has made use of data collected by NASA's Kepler Satellite mission and made publicly accessible
via the NASA Exoplanet Archive, which is operated by the California Institute of Technology, under contract with the National Aeronautics and Space Administration under the Exoplanet Exploration Program.This research has made use of the Exoplanet Orbit Database
and the Exoplanet Data Explorer at exoplanets.org (Han et al. 2014). The simulations described here were performed
on the UCLA Hoffman2 Shared computing cluster.

\newpage

\appendix

\section{Interactions with the Star}
\label{RLO}

In some of the cases discussed in \S~\ref{MultiPerte}, there is only one surviving planet left and this experiences a 
monotonic increase in eccentricity until it hits the star. The fact that this can occur is not a surprise,
if the semi-major axis happens to be at the right location to experience a precession rate resonant with
that of the outer planet system. However, periastra this small result in periastron passages that are 
comparable to or shorter than the timestep in our original simulations. Furthermore, the hybrid symplectic
algorithm of Chambers (1999) may not yield a faithful representation of the orbit in such cases. These
lead to us to question whether the planet truly hits the star or should more simply be regarded as remaining
in a high eccentricity state unless acted on by tides.

\begin{figure}
\includegraphics[width=\columnwidth]{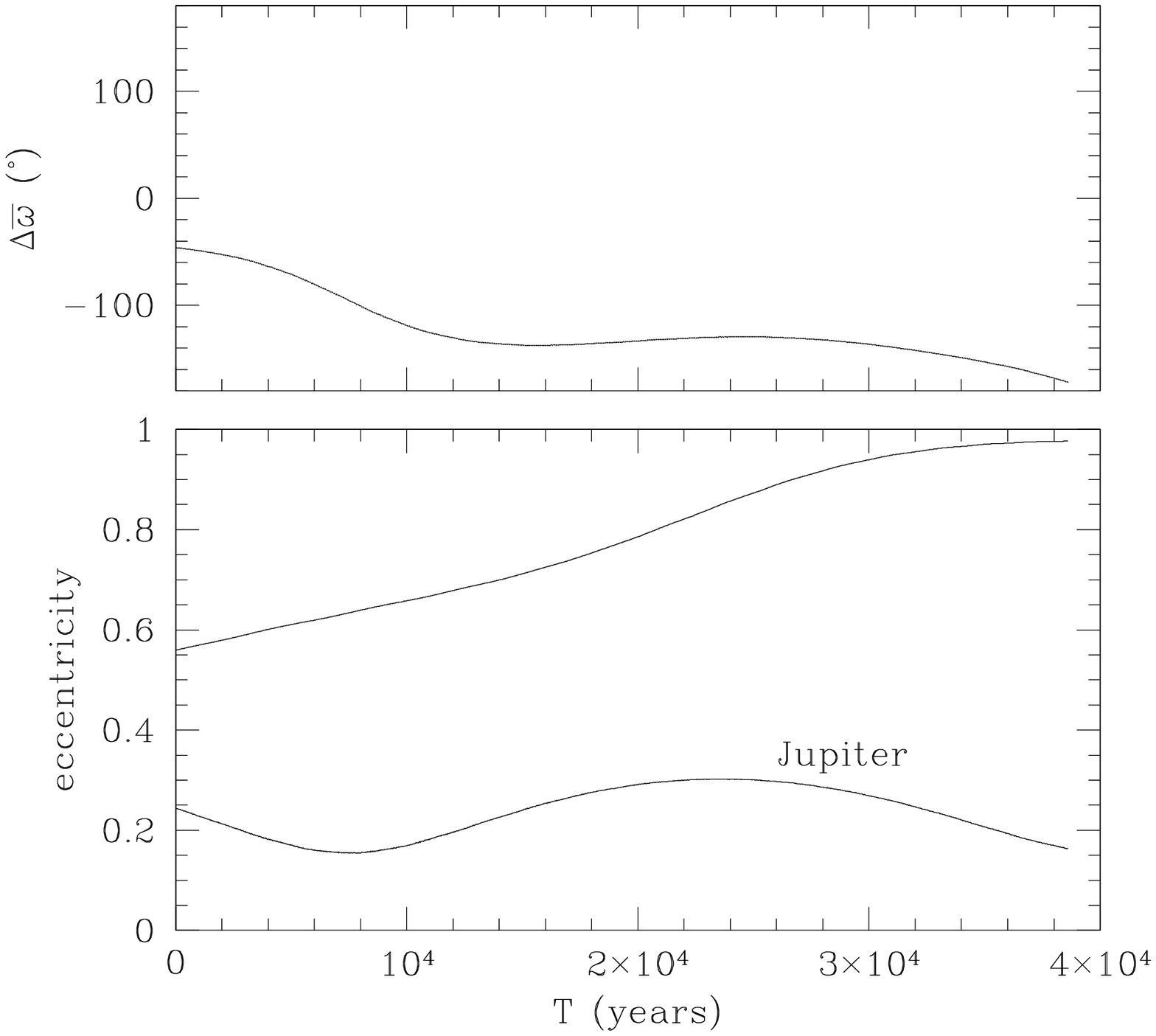}
\caption{The upper panel shows the apsidal alignment of the remaining inner planet with the
Jupiter at 1.25~AU. The fact that this remains negative throughout the course of this integration means
that the eccentricity is continuously pumped. The lower panel shows the resulting eccentricity evolution,
along with that of the perturbing Jupiter. The periastron continues to shrink until the planet hits the star.
\label{fig:Reson2}}
\end{figure}

To answer this question we reran these integrations from the point at which the last planet-planet collision
left a single planet interior to 1~AU. We used a shorter timestep and adopted the Bulirsch-Stoer integrator,
which is a more accurate representation for highly eccentric orbits. The results of one of the examples is
shown in Figure~\ref{fig:Reson2}. We see that this is indeed the result of a secular pumping because the difference
$\Delta \bar{\omega}$ between the longitudes of periastron between the inner planet and the Jupiter does
not circulate over the course of the integration, resulting in a monotonic rise of the eccentricity to the
point of stellar collision. The fact that the integrator is performing faithfully is supported by the fact that
the semi-major axis of the planet does not waver up until the point of stellar collision.

This occurs with non-negligble frequency in cases 
 of eccentricity pumping by a giant planet pair (13/50 cases for Experiments D1 and D3, 2/50 for D2). In each
case, the last planet standing had a semi-major axis in the range $\sim 0.2$--0.3~UA and a moderate eccentricity.
This put it into a configuration of secular resonance and resulted in the pumping of the eccentricity to large
values.

An additional question of interest is whether this planet really does strike the star, or whether it undergoes
Roche-Lobe overflow. The fact that the planet eccentricity rises monotonically means that the planet will have
an opportunity to exceed the Roche limit before it strikes the surface of the star. 
If this were to lead to mass loss and halt the rise to high eccentiricity, this might
provide a pathway to the production of extremely short period planets. 

Such planets have indeed been uncovered in the Kepler data (Rappaport et al., 2012, 2014; Sanchis-Ojeda et al. 2013, 2015; 
Jackson et al. 2013). Sanchis-Ojeda estimate a frequency $\sim 0.8\%$ (averaging their results for G and K stars) per
star, which is about 1\% of all planet hosting stars (comparing to the Petigura et al. estimate). Based on our above
calculation, we could produce substantially more of such objects ($\sim 20\%$ of that particular mode).
Of course, many such objects appear to be evaporating and so may have finite lifetimes, so the survival fraction may
be low.

\label{lastpage}

\end{document}